\documentclass[11pt]{article}
\usepackage{jheppubmodnew}
\pdfoutput=1
\usepackage{comment}
\usepackage{framed}
\usepackage{psfrag}
\usepackage{array}
\usepackage{physics}
\usepackage{bm, bbm, bbold}
\usepackage{amsmath,amsfonts,amssymb, mathrsfs, amsthm}
\usepackage{braket}
\usepackage{cancel}
\usepackage{hyperref}
\usepackage{graphicx}
\usepackage[labelsep=quad]{subcaption}
\usepackage{color,soul}
\usepackage{epsfig}
\usepackage{wrapfig}
\usepackage{environ}
\usepackage{array}
\usepackage{soul}
\usepackage{mathtools}
\usepackage{xspace}

\usepackage{mdframed}

\def \spatinf{\hat{i}^0}
\newcommand{\be}{\begin{equation}}

\newcommand{\ee}{\end{equation}}
\newcommand{\ba}{\begin{align}}
\newcommand{\ea}{\end{align}}
\newcommand{\bs}{\begin{split}}

\newcommand{\momopspat}{\phi_{\spatinf}}
\newcommand{\spatinfann}{\phi_{\spatinf}^{+}}
\newcommand{\spatinfcre}{\phi_{\spatinf}^{-}}
\newcommand{\spatinfop}{\phi_{\spatinf}^{\pm}}

\def\sess\end{split}
\usepackage{float}
\usepackage{simpler-wick}

\newcommand{\w}{{\mathbb{W}}}
\newcommand{\wb}{{\overline{\mathbb{W}}}}
\newcommand{\D}{{\mathbb{D}}}

\newcommand{\TI}{{{\mathbb{T}}}}
\newcommand{\ATI}{{\overline{\mathbb{T}}}}

\def \func{{\cal F}}
\def \orthfunc{{\cal G}}

\def\opspat{{\cal Z}}

\def \scrip{{\cal I}^{+}}
\def \scrim{{\cal I}^{-}}

\def\smone{g}
\def\smtwo{f}

\def \spatinf{\hat{i}^0}

\def\alcut[#1]{{\cal A}_{#1, \epsilon}}
\def\alseg[#1,#2]{{\cal B}_{#1, #2}}

\def\supcharge[#1]{\{#1\}}

\def\projsupeig[#1]{{\cal P}_{{\ell, m}}[{#1}]}
\def\transop[#1, #2]{T_{\{#1\}, \{#2\}}}
\def\supket[#1]{|\{#1\} \rangle}
\def\supbra[#1]{\langle \{#1\} | }

\def\hilbzerosect[#1]{{\cal H}^0_{\{#1\}}}
\def\hilbsect[#1]{{\cal H}_{\{#1\}}}
\def\hilbmass[#1]{{\cal H}^m}

\def\Or[#1]{\text{O}\!\left(#1\right)}
\newcommand{\lin}{{_\text{in} \langle}}
\newcommand{\rin}{{\rangle_{\text{in}}}}
\newcommand{\rinp}{{\rangle'_{\text{in}}}}
\newcommand{\lout}{{_{\text{out}} \langle}}
\newcommand{\rout}{{\rangle_{\text{out}}}}

\def\smrf{{\mathcal{K}}}

\def\dds{d \mu_{X}}

\def\phipert[#1]{\delta \phi^{(#1)}}
\def\opspatpert[#1]{\delta \opspat^{(#1)}}

\def\phitailsub[#1]{\widetilde{\phi}^{(#1)}}
\def\phinupert[#1]{\delta \phi_{\nu}^{(#1)}}

\newcommand{\inim}{{\em in}\xspace}
\newcommand{\outip}{{\em out}\xspace}

\begin{document} 
\title{Interacting Fields at Spatial Infinity}
\author{Anupam A. H,}
\author{P.V. Athira,}
\author{Priyadarshi Paul}
\author{and  Suvrat Raju}
\emailAdd{anupam.ah@icts.res.in}
\emailAdd{athira.pv@icts.res.in}
\emailAdd{priyadarshi.paul@icts.res.in}
\emailAdd{suvrat@icts.res.in}

\affiliation{International Centre for Theoretical Sciences, Tata Institute of Fundamental Research, Shivakote,
Bengaluru 560089, India.}

\abstract{We study the properties of massive fields extrapolated to the blowup of spatial infinity ($\spatinf$), extending the program initiated in \texttt{arXiv:2207.06406}. In the free theory, we find an explicit representation of boundary two-point functions and boundary to bulk two-point functions, and also present an HKLL-type reconstruction formula for local bulk operators in terms of smeared boundary operators.  We study interacting Wightman correlators and find that, generically, interacting massive fields decay slower than free fields as one approaches $\spatinf$. We propose that meaningful correlators at $\spatinf$ can be obtained through an LSZ-like prescription that isolates the on-shell part of bulk Wightman correlators before extrapolating them to $\spatinf$. We show that a natural basis for operators at $\spatinf$, defined via this prescription, is given by the average of ``in'' and ``out'' operators defined at $i^-$ and $i^+$ respectively. Therefore, correlators at $\spatinf$ and cross correlators between $\spatinf$, $i^-$ and $i^+$ can be represented within the class of asymptotic observables studied by Caron-Huot et al. in \texttt{arXiv:2308.02125}. We present several sample calculations.}

\setcounter{tocdepth}{1}
\maketitle
\noindent
\flushbottom
\section{Introduction}
The authors of \cite{Laddha:2022nmj} introduced a new class of asymptotic operators by extrapolating free massive fields to a blowup of spatial infinity.  This blowup, which we denote by $\spatinf$,  corresponds to the locus of points at infinite proper distance from a chosen origin. It is natural to think of $\spatinf$ as a copy of dS$_3$. The extrapolation procedure described in \cite{Laddha:2022nmj} is similar, in spirit, to the procedure used to define boundary operators as limits of bulk operators in AdS/CFT  \cite{Witten:1998qj,Gubser:1998bc,Banks:1998dd}.

The boundary algebra at $\spatinf$ is delicate. This can be traced back to the fact that operators on $\spatinf$ are obtained by rescaling bulk fields with a factor that grows exponentially with the proper distance. This is in contrast to the AdS case, where the rescaling factor grows only as a power law.  For free fields, \cite{Laddha:2022nmj} showed that operators on $\spatinf$ must be smeared with a restricted class of smearing functions to obtain observables with finite fluctuations. Said another way, boundary operators on $\spatinf$ are distributions whose domain comprises a restricted class of test functions.

This paper's main aim is to define a boundary algebra at $\spatinf$ for interacting bulk theories. However, we first extend the results of \cite{Laddha:2022nmj}. We show that the allowed class of smearing functions described there corresponds to functions that are analytic under a complexification of the boundary dS$_3$.  This observation allows us to obtain an explicit formula for the two-point function of boundary operators in free-field theory. It also allows us to compute an HKLL-type transfer function \cite{Hamilton:2005ju, Hamilton:2006az, Hamilton:2006fh, Hamilton:2007wj,Kabat:2011rz} that, in the free theory,  can be used to reconstruct fields on all of Minkowski space given their boundary values on $\spatinf$. 

We then turn to a study of interacting Wightman correlators at $\spatinf$. Wightman correlators are those where one picks an ordering of operators by hand. We study Wightman correlators instead of time-ordered correlators because of the feature explained above: even in free-field theory,  good boundary operators must be smeared with analytic functions on $\spatinf$.   Time-ordering would introduce additional step functions in the smearing functions that are disallowed by analyticity.  

We show that naively applying the procedure of \cite{Laddha:2022nmj} to interacting fields leads to an ill-defined algebra at $\spatinf$. This is because the falloff of interacting fields at large distance can differ from free fields. However, we propose that a natural, field-redefinition-independent piece of an interacting correlator, with the correct falloff, can be extracted by generalizing the LSZ formula to Wightman correlators.  

The standard LSZ formula tells us that the S-matrix is the residue of the pole in the bulk time-ordered correlator. Analogously, we focus on the term in the bulk Wightman correlator that contains mass-shell delta functions for each external momentum. We demonstrate that this term is independent of field redefinitions and exhibits the appropriate falloff for extrapolation to $\spatinf$.

We show that a natural basis of operators at $\spatinf$ can be constructed by taking the average of ``\inim'' fields and ``\outip'' fields, which are defined at past infinity ($i^-$) and future infinity ($i^+$) respectively. This identifies correlators at $\spatinf$ as a subclass of the asymptotic observables that were recently studied in \cite{Caron-Huot:2023vxl}. It also implies that all correlators at $\spatinf$ can be written in terms of polynomials of the S-matrix and its conjugate.  So a knowledge of the S-matrix completely fixes the algebra at $\spatinf$. 

In perturbation theory, it is straightforward, albeit tedious, to compute Wightman correlators and implement our on-shell truncation procedure. We provide several perturbative examples in section \ref{secsample}.

The usual $S$-matrix formalism refers to field operators extrapolated to $i^-$ and $i^+$ where geodesics of massive particles start and end.  The reader might wonder why we focus on extrapolating massive fields to $\spatinf$. As explained in \cite{Laddha:2022nmj}, the motivation for this study comes from work on the ``holography of information''. We would like to advance the idea that, in a theory of quantum gravity in asymptotically flat space, observables in the bulk of a Cauchy slice can be equated to observables near its boundary. 

In flat space, this idea can be made precise for massless fields that can be extrapolated to null infinity \cite{Laddha:2020kvp}. (For related work in flat space and AdS see \cite{Chowdhury:2020hse,Raju:2020smc,Chowdhury:2021nxw,Raju:2021lwh,Chakravarty:2023cll,deMelloKoch:2022sul,deMelloKoch:2024juz}  and for an implementation of this idea in dS see \cite{Chakraborty:2023los,Chakraborty:2023yed}.) However, there does not appear to be any straightforward method of studying massive fields on null infinity \cite{Helfer:1993hv}. The blowup of spatial infinity provides a natural boundary for flat space and so it is worth investigating whether massive fields can be extrapolated to spatial infinity. We refer the reader to \cite{Marolf:2006bk} for earlier work on extrapolating massive fields to the blowup of spatial infinity, which was also motivated by flat-space holography.

While we eventually hope to apply this formalism to quantum gravity, we emphasize that the results in this paper are restricted to quantum field theory in pure-Minkowski space.

It would clearly be of interest to relate the formalism developed here to the study of celestial holography \cite{Fan:2020xjj,Albayrak:2020saa,Law:2020xcf,Raclariu:2021zjz,McLoughlin:2022ljp,Pasterski:2021raf,Donnay:2022aba,Donnay:2021wrk}. Perhaps the study of spatial infinity provides a method of unifying work on the holography of information with work on celestial holography. It is interesting that one of the precursor-papers for celestial holography \cite{deBoer:2003vf} also focused on exploring flat-space quantum field theory in dS and AdS slicings.

An overview of this paper is as follows. In section \ref{prelim}, we review the setup of \cite{Laddha:2022nmj}. In section \ref{sectwoptfree}, we extend the free-field results of \cite{Laddha:2022nmj}. In section  \ref{secbareinteracting}, we initiate an analysis of interacting correlators at spatial infinity and show why a naive extension of the free-field extrapolation procedure fails in the interacting theory. In section \ref{seconshellinteract}, we describe an extension of the LSZ-prescription to Wightman correlators.  This leads to an interesting algebra of interacting operators at spatial infinity. In section \ref{secspatinfinout} we relate this algebra to the standard \inim and \outip algebras. In section \ref{secsample}, we provide several examples of correlation functions involving operators at $\spatinf$ and also correlators that involve processes with insertions at $i^{-}, i^{+}$ and $\spatinf$. Appendix \ref{reviewfeynman} provides a detailed review of perturbation theory for Wightman functions and Appendix \ref{details_of_comp} provides details of the computations whose results are presented in section \ref{secsample}.

Section \ref{secsummary} contains a summary of all the main results obtained in this paper. The reader might want to start the paper by consulting that section. 

\section{Review and setup} \label{prelim}
In this section, we provide a lightning review of the setup and main results of  \cite{Laddha:2022nmj}. 

\subsection{The blowup of spatial infinity, $\spatinf$}
In the usual Penrose diagram, spatial infinity is a singular point. However, it can be ``blown up'' to yield a copy of dS$_3$ \cite{Ashtekar:1978zz,Ashtekar:1991vb} as we now review.\footnote{We refer the reader to \cite{Friedrich:1999ax,Friedrich:1999wk,Friedrich:2002ru,Friedrich:2006km} for an alternate construction that is geometrically equivalent \cite{Mohamed:2021yzf}.}

Starting with the Minkowski metric,
\be
ds^2 = \eta_{\mu \nu } dx^{\mu}dx^{\nu} = - dt^2 + d r^2 + r^2 d \Omega^2~,
\ee
we move to the coordinates
\be 
t = \rho \sinh{\tau }, \quad r = \rho \cosh \tau,
\ee
outside the light cone of an arbitrarily chosen origin.
\begin{figure}[h]
    \centering
\includegraphics{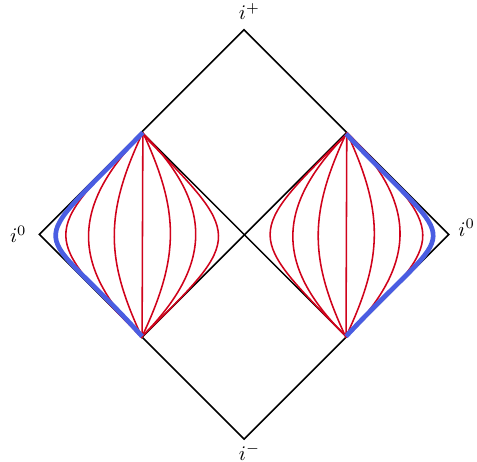}
    \caption{de Sitter slicing of flat spacetime which covers the region where $\abs{t}< r$.  Constant $\rho$ hypersurfaces are dS$_3$ slices, indicated by red lines. $\spatinf$ refers to the asymptotic dS$_3$ slice indicated in blue.} 
    \label{fig:ds_slices}
\end{figure}

The metric in this coordinate system is 
\be
ds^2 = d \rho^2 + \rho^2 \underbrace{( - d \tau^2 + \cosh^2 \tau ~d \Omega^2 )}_{dS_3 \text{ metric }}~.
\ee
It is evident that constant $\rho$ hypersurfaces are copies of dS$_3$. (See Figure \ref{fig:ds_slices}.) It is convenient to parameterize them using the embedding coordinates $X^{\mu} = \tfrac{x^{\mu}}{\rho}$ that satisfy $\eta_{\mu \nu}X^{\mu}X^{\nu} =1~$. 
Explicitly $X^{\mu}$ can be written in terms of $\{\tau , \Omega\}$ as
\be\label{para0}
    X^{\mu} = \{ \sinh \tau\,,\,\cosh \tau \,\Omega_x\,,\, \cosh \tau \,\Omega_y\,,\,\cosh \tau \,\Omega_z\},
\ee
where $(\Omega_x, \Omega_y, \Omega_z)$ specifies a unit vector on $S^2$.

\paragraph{\bf Coordinates.} A bulk point can be specified by a four-vector, $x$, or using $(t, \vec{x})$,  or in terms of $(t, r, \Omega)$. Alternately, one can use the coordinates $(\rho, X)$. The de Sitter coordinate $X$ can, in turn, be written in terms of $\tau, \Omega$.  These are related as above, and we will freely switch between these coordinate systems below. 

\paragraph{Blowup of spatial infinity.}
In this paper, following \cite{Laddha:2022nmj},  the blowup of spatial infinity, $\spatinf$, is defined as the dS$_3$ slice obtained in the limit $\rho \rightarrow \infty$ i.e.  it is the dS$_3$ slice that consists of points at an infinite proper distance from a chosen origin. 

In the gravitational literature, this limit is taken with the additional understanding that $\tau$ is {\em not} taken to infinity as  $\rho$ is taken to infinity.  However, in our analysis we will sometimes take $\rho, \tau$ to infinity simultaneously. So $\spatinf$ should be understood as the asymptotic dS$_3$ that interpolates between past and future null infinity. 

Our choice of an origin breaks translational invariance. However, asymptotic operators on $\spatinf$ transform nicely under translations  \cite{Laddha:2022nmj}. 

\subsection{Extrapolation procedure for free fields}
The massive free field equation in $\{ \rho , X\}$ coordinates
\be
( \Box - m^2 )\phi (\rho, X) =0,
\ee
is solved by 
\be
\phi(\rho, X) =  \int_0^{\infty} d \nu \frac{K_{i \nu}(m \rho)}{\rho} \phi _{\nu}(X)~,
\ee
where $\phi_{\nu}(X)$ obeys the massive de Sitter wave equation \cite{Bros:1995js}
\be
(\Box_{dS_3}- (1+ \nu ^2))\phi_{\nu} (X) =0~.
\ee

Therefore, a single massive free field in Minkowski space breaks up into an infinite number of massive de Sitter free fields.  The field $\phi_{\nu}(X)$ lives in the principal-series representation of dS$_3$.  However, these component fields will soon be reassembled into a boundary operator, and the individual components will not play any significant role in the discussion.

In the large $\rho$ limit ($m \rho \gg \nu$),
\be
\frac{K_{i \nu}(m \rho)}{\rho} \xrightarrow[\rho \rightarrow \infty]{ } \sqrt{\frac{\pi}{2}} \sqrt{\frac{1}{m \rho}} \frac{e^{- m \rho}}{\rho }~,
\ee
and so the radial function has a universal form at large $\rho$ independent of $\nu$  \cite{dunster1990bessel}.
With 
\be
D(\rho) = \sqrt{\frac{2}{\pi}}\rho\,\sqrt{m \rho}  \, {e^{m \rho}}~,
\ee
we define boundary operators in the free theory through
\be
\label{ptwiseextrapolate}
\opspat(X) \equiv \lim_{\rho \rightarrow \infty} D(\rho) \phi(\rho,X) = \int d \nu \, \phi_{\nu}(X)~.
\ee
This limit is subtle. Even in the free theory, as we review below, the extrapolated fields $\opspat(X)$ should be viewed as operator-valued distributions whose domain is restricted to a specific class of functions. In the interacting theory, it is additionally necessary to first extract the on-shell part of the bulk field before taking the boundary limit.

\subsection{Smearing operators on $\spatinf$}
In \cite{Laddha:2022nmj}, it was observed the two-point function $\langle \Omega| \opspat(X_1) \opspat(X_2) | \Omega \rangle$ was not well defined in the free theory for general values of $X_1, X_2$. This can be understood by examining the bulk two-point function $\langle \Omega | \phi(\rho, X_1) \phi(\rho, X_2) | \Omega \rangle$. For generic positions of $X_1, X_2$, the distance between the two bulk points 
\be
|\rho X_1 - \rho X_2| = \rho(2 - 2 X_1 \cdot X_2)^{1 \over 2},
\ee
is different from $2 \rho$. Therefore scaling the bulk correlator with a factor of $D(\rho)^2$ as one takes $\rho \rightarrow \infty$ produces either $0$ or $\infty$ except for special configurations of $X_1, X_2$.

To address this problem, \cite{Laddha:2022nmj} proposed that one should instead consider the smeared field 
\be
   \opspat(\smone) = \int  \dds \smone(X) \opspat(X)~,
\ee
where, $\dds \equiv \,\cosh^{2}\tau d\tau d\Omega$ for a restricted class of smearing functions $\smone(X)$. More precisely, in the {\em free theory}, given two smearing functions $\smone(X), \smtwo(X)$, we define
\be
\label{freeextrapolate}
\langle \Omega| \opspat(\smone) \opspat(\smtwo) |\Omega \rangle = \lim_{\rho \rightarrow \infty} D(\rho)^2 \int {\dds}_1 {\dds}_2 \langle \Omega| \phi(\rho, X_1) \phi(\rho, X_2) |\Omega \rangle \smone(X_1) \smtwo(X_2), 
\ee
where the large $\rho$ limit is taken {\em after} we perform the smearing in $X_1$ and $X_2$.  

We now describe the restrictions on smearing functions that were found in  \cite{Laddha:2022nmj}. The formulas that follow might seem complicated but we reassure the reader that we will soon reinterpret them in simple terms.  Write the smearing function in terms of $c$-numbers $\smone_{\nu, \ell}$ using
\be
\smone(X) = \sum_{\ell} \int_{0}^{\infty} d \nu \,  \smone_{\nu, \ell}  \, \orthfunc_{\nu,\ell}(\tau) \, Y^*_{\ell}(\Omega) + \text{h.c.} ,  
\ee
where
\be
\orthfunc_{\nu, \ell}(\tau) = {1+e^{-2 \pi \nu}  \over 2 \pi ( 1 - e^{-2 \pi \nu})^2 } \, \func^*_{\nu, \ell} (\tau) - {c^{*}_{\nu,\ell} \over \pi (1 - e^{-2 \pi \nu})^2 } \, \func_{\nu, \ell} (\tau), \ee
with
\be
\label{modefunc}
\begin{split}
\func_{\nu,\ell}(\tau) = {2 e^{\ell \tau} \over  (2 \cosh \tau)^{\ell+1}} \Big[&e^{-i \nu \tau}  \, _2F_1\left(i \nu - \ell, -\ell, 1 + i \nu, -e^{-2 \tau} \right) 
\\ &+ c_{\nu, \ell} e^{i \nu \tau} \, _2F_1\left(-i \nu - \ell, -\ell, 1 - i \nu, -e^{-2 \tau} \right) \Big],
\end{split}
\ee
and
\be
c_{\nu, \ell} =-e^{-\pi  \nu } \frac{ \Gamma (i \nu +1) \Gamma (\ell-i \nu +1)}{\Gamma (1-i \nu ) \Gamma (\ell+i \nu +1)} .
\ee
Here $Y_{\ell}$ are the standard spherical harmonics. Both $Y_{\ell}$ and the $c$-number coefficients, $\smone_{\nu, \ell}$ depend on a magnetic quantum number, $m$, although this dependence is suppressed to lighten the notation. The function $\orthfunc_{\nu, \ell}$ is independent of the magnetic quantum number.  We will also use the notation
\be
Y_{-\ell}(\Omega) \equiv Y_{\ell}(\Omega)^*.
\ee

Then it was shown in \cite{Laddha:2022nmj} that the two-point function is well-defined provided
\be\label{smearingnufalloff}
		\begin{split}
		\smone_{\nu,\ell} \ &\xrightarrow[\nu \rightarrow \pm \infty]{} \  \, \frac{e^{- \frac{\pi|\nu|}{2}}}{\nu^{\alpha}} , \qquad \alpha\, >\, \frac{1}{2}\, \forall\, \ell,\\
		\sum_{m} |\smone_{\nu,\ell}|^2 \ \ &\xrightarrow[\ell \rightarrow \infty]{}  \  \ \ \, {1\over {\ell}^{\beta} }, \qquad \qquad \beta > 1 \,  \forall \, \nu.
		\end{split}
	\ee 
The same condition also makes the fluctuations of a boundary observable in the vacuum, $\langle \Omega | \opspat(\smone) \opspat(\smone) |\Omega \rangle$, finite.

\section{Revisiting the free theory \label{sectwoptfree}}
In this section, we revisit and extend the free-field results reviewed above. First, we will show that the technical-looking restriction placed on smearing functions in \cite{Laddha:2022nmj} has a simple interpretation in terms of its analytic properties. We will use this observation to obtain compact expressions for two-point functions and a HKLL-type kernel \cite{Hamilton:2005ju, Hamilton:2006az, Hamilton:2006fh, Hamilton:2007wj,Kabat:2011rz} for bulk reconstruction from the boundary.

\subsection{Analytic properties of smearing functions}

Recalling the hypergeometric series
\be
_2 F^1 (a,b,c,x) = \sum_{n=0}^{\infty} {(a)_n (b)_n \over (c)_n} {x^n \over n!},
\ee
we see that the hypergeometric functions that appear in \eqref{modefunc} reduce to finite polynomials because $\ell$ must be an integer.  Therefore the mode functions are just oscillating exponentials multiplied by a function of $\tau$ that dies off exponentially as $\tau \rightarrow \pm \infty$. 

Now consider
\be
\smone_{\ell}(\tau) = \int \smone(\tau, \Omega) Y_{\ell}(\Omega) d \Omega.
\ee
Note that
\be
\smone(\tau, \Omega) = \sum_{\ell} \smone_{\ell}(\tau) Y_{\ell}^*(\Omega).
\ee
The sum over $\ell$ runs over all spherical harmonics but since we have suppressed the magnetic quantum number and $\ell$ is always positive we will use the notation
\be
\smone_{-\ell}(\tau)=\int \smone(\tau, \Omega)Y_{\ell}^*(\Omega) d \Omega.
\ee

The condition on the coefficients \eqref{smearingnufalloff} --- which requires them to fall off exponentially with $\nu$ --- immediately implies that 
\be
\label{analdomain}
\smone_{\ell}(\tau)~\text{is~analytic~for}~\Im(\tau) \in (-{\pi \over 2}, {\pi \over 2}), \quad \forall \ell.
\ee
 Except for the value $\Re(\tau) = 0$,  we expect $\smone_{\ell}(\tau)$ to be continuous as we approach the lines $\Im(\tau) = \pm {\pi \over 2}$ from within the domain of analyticity.

At  $\Im(\tau) = \pm {\pi \over 2}$, we see that \eqref{smearingnufalloff} allows $\smone_{\ell}(\tau)$ to have a pole of order $\ell + 1$ at $\Re(\tau) = 0$. This pole can arise from the vanishing of $\cosh(\tau)$ in the denominator of \eqref{modefunc}.\footnote{Recall that $\orthfunc$ is a linear combination of $\func$ and $\func^*$; one of these is regular at $\Im(\tau) = {\pi \over 2}$ and the other at $\Im(\tau) = -{\pi \over 2}$. But $\orthfunc$ has a pole in both cases.}  However, a careful consideration of the mode functions shows that we have
\be
\label{poleproperty}
\lim_{\xi \rightarrow 0} \smone_{\ell}(\xi \pm {i \pi \over 2}) + (-1)^{\ell} \smone_{\ell}(\xi \pm {i \pi \over 2}) = 0.
\ee
We will see that this is the combination that appears below. 

The Lorentz invariant version of the analyticity criterion is as follows. Consider an extension of $X$ with a parameter, $\beta$, and another vector $Y$ through
\be
\label{xextn}
X \rightarrow X \cos \beta  + i Y \sin \beta,
\ee
where $Y$ satisfies
\be
Y\cdot X = 0; \qquad Y \cdot Y = -1.
\ee
Then we demand that the smearing function, $\smone(X)$,  be analytic under extensions of the form \eqref{xextn} in the range
\be
\beta \in (-{\pi \over 2}, {\pi \over 2}).
\ee
Some smearing functions that were studied in \cite{Laddha:2022nmj} could have branch cuts as one approaches $\beta = \pm {\pi \over 2}$. So it is important to take this limit from within the domain of analyticity.

\subsection{Extrapolated smeared plane waves \label{subsecextrapplane}}
We now derive a preliminary mathematical result that will be very useful in what follows. Let $\vec{k}$ be a three vector and define the four vector
\be
k=(\omega_k, \vec{k}),
\ee
with $\omega_k = \sqrt{\vec{k}^2 + m^2}$.  We also define $\tau_0$ via
\be
m \cosh \tau_0 = \omega_k.
\ee
We will derive an expression for the smeared extrapolated plane waves
\be
\begin{split}
\widetilde{\smone}^{-}(\vec{k}) \equiv &\lim_{\rho \rightarrow \infty} D(\rho) \int e^{-i \rho k \cdot X} \smone(X) \dds; \\
\widetilde{\smone}^{+}(\vec{k}) \equiv &\lim_{\rho \rightarrow \infty} D(\rho) \int e^{i \rho k \cdot X} \smone(X) \dds,
\end{split}
\ee
where $\smone(X)$ is a smearing function with the properties above. 

We start by writing the spatial exponential as  
\be
e^{\pm i \vec{k} \cdot \vec{x}} =  4 \pi \sum_{\ell} i^{\pm \ell} j_{\ell}(|\vec{k}| r) Y_{\ell}^*(\hat{k}) Y_{\ell}(\Omega),
\ee
and so
\be
\int e^{\pm i \vec{k} \cdot \vec{x}} Y^*_{\ell}(\Omega) d^2 \hat{x} = 4 \pi  \sum_{\ell} i^{\pm \ell} j_{\ell}(|\vec{k}| r) Y^*_{\ell}(\hat{k}).
\ee
Using the asymptotic behaviour of the spherical Bessel function 
\be
j_{\ell}(|\vec{k}| r ) \underset{r \rightarrow \infty}{\longrightarrow} {1 \over 2 |\vec{k}| r} \left(i^{-(\ell+1)} e^{i |\vec{k}| r} + i^{(\ell+1)} e^{-i |\vec{k}| r}\right),
\ee
noting that the dS measure is $\dds = \cosh^2 \tau_{} d \tau_{} d^2 \Omega$, and doing the trivial integral over $\Omega$ using the orthogonality of the spherical harmonics  we find that
\be
\begin{split}
\widetilde{\smone}^{-}(\vec{k}) = \lim_{\rho \rightarrow \infty} {2 \pi i  D(\rho) \over m \rho \sinh \tau_0} \sum_{\ell} \int \big((-1)^{\ell+1} e^{i m \rho \sinh(\tau + \tau_0)}  + e^{i m \rho \sinh(\tau - \tau_0)} \big) \smone_{\ell}(\tau) Y_{\ell}^*(\hat{k}) \cosh \tau d \tau, 
\end{split}
\ee
and likewise
\be
\widetilde{\smone}^{+}(\vec{k}) =  \lim_{\rho \rightarrow \infty} {-2 \pi i  D(\rho) \over m \rho \sinh \tau_0} \sum_{\ell} \int  \big( (-1)^{\ell+1} e^{-i m \rho \sinh(\tau + \tau_0)}  +  e^{-i m \rho \sinh(\tau - \tau_0)} \big) \smone_{\ell}(\tau) Y_{\ell}^*(\hat{k})\cosh \tau d \tau.
\ee

Since $\rho$ is large, we can perform the integrals over $\tau_{}$  using a saddle point approximation. Each exponential factor has two saddle points when $\cosh(\tau_{} \pm \tau_0) = 0$.  Both of these are off the original contour of integration. However, the contour of integration can be deformed to be a contour of constant phase that passes through one of these saddles.

The imaginary part of the exponents that appear in the integrand, after accounting for the explicit factor of $i$ and noting that $\tau_0$ is real, is proportional to
\be
\Re(\sinh(\tau_{} \pm \tau_0)) = \cos(\Im(\tau_{})) \sinh(\Re(\tau_{}) \pm \tau_0).
\ee
Therefore both choices, $\Im(\tau) = \pm {\pi \over 2}$ correspond to contours of constant phase. But only one choice leads to a convergent integral as indicated in the table below.
\begin{table}[H]
\centering
\setlength{\extrarowheight}{2pt}
\begin{tabular}{|c|c|}
\hline
{Term} & {Contour}  \\ [0.5ex]
\hline
$e^{i m \rho \sinh(\tau_1 - \tau_0)}$  &  $\text{Im}(\tau_1) = {\pi \over
2}$\\
[0.5ex]
\hline
$e^{i m \rho \sinh(\tau_1 + \tau_0)} $& $\text{Im}(\tau_1) = {\pi \over 2}
$ \\
[0.5ex]
\hline
$e^{-i m \rho \sinh(\tau_1 - \tau_0)}$ & $\text{Im}(\tau_1) = -{\pi \over
2}$ \\
[0.5ex]
\hline
$e^{-i m \rho \sinh(\tau_1 + \tau_0)} $& $\text{Im}(\tau_1) = -{\pi \over
2}$
\\
[0.5ex]
\hline
\end{tabular}
\label{}
\end{table}
If there are poles on these contours, we avoid them by going through the domain of analyticity. In all cases, the contour passes through the exponentially decaying saddle point that provides the dominant contribution to the integral.  This is indicated in the table below. 

\begin{table}[H]
\centering
\setlength{\extrarowheight}{2pt}
\begin{tabular}{|c|c|c|}
\hline
{Term} & {Saddles} &{Contributing~saddle} \\ [0.5ex]
\hline
$e^{i m \rho \sinh(\tau_1 - \tau_0)} $ &$ \tau_1 = \tau_0 \pm i {\pi \over
2}$ &$ \tau_1 = \tau_0 + i {\pi \over 2}$ \\
[0.5ex]
\hline
$e^{i m \rho \sinh(\tau_1 + \tau_0)} $&$ \tau_1 = -\tau_0 \pm i {\pi \over
2} $&$ \tau_1 = -\tau_0 + i {\pi \over 2} $\\
[0.5ex]
\hline
$ e^{-i m \rho \sinh(\tau_1 - \tau_0)}$ &$ \tau_1 = \tau_0 \pm i {\pi \over
2} $&$ \tau_1 = \tau_0 - i {\pi \over 2} $\\
[0.5ex]
\hline
$e^{-i m \rho \sinh(\tau_1 + \tau_0)}$ &$ \tau_1 = -\tau_0 \pm i {\pi \over
2}$ &$ \tau_1 = -\tau_0 - i {\pi \over 2} $ \\
[0.5ex]
\hline
\end{tabular}
\label{}
\end{table}

Picking up the contribution from the correct saddle and including the Gaussian fluctuations, we find that all the $\rho$-dependence in the factor of $D(\rho)$ cancels off nicely and we are left with the following final results.
\be
\label{fplustrans}
\widetilde{\smone}^{-}(\vec{k})= -{4 \pi \over m} \sum_{\ell}   \left( \smone_{\ell}(\tau_0 + i {\pi \over 2}) + (-1)^{\ell} \smone_{\ell}(-\tau_0 + i {\pi \over 2}) \right) Y^*_{\ell}(\hat{k}),
\ee
and,
\be
\label{fminustrans}
\widetilde{\smone}^{+}(\vec{k}) = -{4 \pi \over m} \sum_{\ell}  \left(\smone_{\ell}(\tau_0 - i {\pi \over 2}) + (-1)^{\ell} \smone_{\ell}(-\tau_0 - i {\pi \over 2}) \right) Y^*_{\ell}(\hat{k}).
\ee
Equations \eqref{fminustrans} and \eqref{fplustrans} are simple but important mathematical results that we will use multiple times below.

\subsection{Explicit form of two-point functions}
Some sample two-point correlators were already computed in \cite{Laddha:2022nmj}. However, the formula above provides a more compact expression for two-point functions corresponding to general smearing functions.

We start with the expression for the two-point bulk Wightman function
\be
\label{bulktwopt}
\langle \Omega | \phi(x_1) \phi(x_2) | \Omega \rangle = \int e^{i k \cdot (x_1 - x_2)} (2 \pi) \delta(k^2 +  m^2) \theta(k^0) {d^4 k \over (2 \pi)^4}.
\ee
We will use this to determine the boundary two-point function and the bulk-boundary two-point function.

\subsubsection{Boundary two-point function}
The boundary two-point function is
\be
\langle \Omega | \opspat(\smone) \opspat(\smtwo) | \Omega \rangle =  \lim_{\rho \rightarrow \infty} D(\rho)^2 \int \langle \Omega | \phi(\rho, X_1) \phi(\rho, X_2) | \Omega \rangle \smone(X_1) \smtwo(X_2){\dds}_1 {\dds}_2.
\ee
Using the results above we see that
\be
\langle \Omega | \opspat(\smone) \opspat(\smtwo) | \Omega \rangle = \int \widetilde{\smone}^{+}(\vec{k}) \widetilde{\smtwo}^{-}(\vec{k}) {d^3 \vec{k} \over (2 \pi)^3 2 \omega_k}.
\ee
With a change of variables to $|\vec{k}| = m \sinh \tau_0$ this can be written as
\be
\label{bdrybdry}
\begin{split}
\langle \Omega | \opspat(\smone) \opspat(\smtwo) | \Omega  \rangle = {1 \over  \pi}  \sum_{\ell} \int_0^{\infty} d \tau_0 \sinh^2 \tau_0 &\left(\smone_{\ell}(\tau_0 - i {\pi \over 2}) + (-1)^{\ell} \smone_{\ell}(-\tau_0 - i {\pi \over 2}) \right) \\ &\times \left(\smtwo_{-{\ell}}(\tau_0 + i {\pi \over 2}) + (-1)^{\ell} \smtwo_{-\ell}(-\tau_0 + i {\pi \over 2})  \right).
\end{split}
\ee
We can change the range of the integral over $\tau_0$ to write
\be
\begin{split}
\langle \Omega |  \opspat(\smone) \opspat(\smtwo) | \Omega \rangle = &{1 \over \pi} \sum_{\ell}  \int_{-\infty}^{\infty} d \tau_0 \sinh^2 \tau_0 \\ & \times \left[\smone_{\ell}(\tau_0 - i {\pi \over 2}) \smtwo_{-{\ell}}(\tau_0 + i {\pi \over 2})  + (-1)^{\ell} \smone_{\ell}(-\tau_0 - i {\pi \over 2})   \smtwo_{-{\ell}}(\tau_0 + i {\pi \over 2}) \right].
\end{split}
\ee

\subsection{Boundary-bulk two-point function}
We can also obtain the correlator when one point is at the boundary and the other point is in the bulk. To do this, we simply repeat the calculation above while taking only one-point to infinity.
\be
\langle \Omega | \opspat(\smone) \phi(\rho_2, X_2) | \Omega \rangle =  \lim_{\rho_1 \rightarrow \infty} D(\rho_1) \int \langle \Omega | \phi(\rho_1, X_1) \phi(\rho_2, X_2) | \Omega \rangle \smone(X_1) {\dds}_1.
\ee
Using the analysis above, 
\be
\langle \Omega | \opspat(\smone) \phi(\rho_2, X_2) |\Omega \rangle = \int \widetilde{\smone}^{+}(\vec{k}) e^{-i k \cdot x_2} {d^3 \vec{k} \over (2 \pi)^3 2 \omega_{k}}.
\ee

Using the same decomposition for the smearing function $\smone$ as above and doing the angular integrals we find that
\be
\label{bulkbdry}
\begin{split}
\langle \Omega | \opspat(\smone) \phi(\rho, X_2) | \Omega \rangle = {-m \over \pi} \sum_{\ell} \int_0^{\infty} &d \tau_0 \big(\smone_{\ell}(\tau_0 - i {\pi \over 2}) + (-1)^{\ell} \smone_{\ell}(-\tau_0 - i {\pi \over 2}) \big) \\ &\times  e^{i m \cosh \tau_0 t_2} j_{\ell}(m \sinh \tau_0 r_2) i^{-\ell} Y_{\ell}^*(\Omega_2)  \sinh^2 \tau_0.
\end{split}
\ee
With the other ordering one has
\be
\label{bulkbdryconj}
\begin{split}
\langle  \Omega | \phi(\rho, X_2) \opspat(\smone) | \Omega  \rangle = {-m \over \pi} \sum_{\ell} \int_0^{\infty} &d \tau_0 \big(\smone_{-\ell}(\tau_0 + i {\pi \over 2}) + (-1)^{\ell} \smone_{-\ell}(-\tau_0 + i {\pi \over 2}) \big) \\ &\times  e^{-i m \cosh \tau_0 t_2} j_{\ell}(m \sinh \tau_0 r_2) i^{\ell} Y_{\ell}(\Omega_2)  \sinh^2 \tau_0.
\end{split}
\ee

\subsection{HKLL-type smearing function}
In this subsection we show that it is possible to use the results of the previous section to determine a HKLL-type smearing function \cite{Hamilton:2005ju, Hamilton:2006az, Hamilton:2006fh, Hamilton:2007wj,Kabat:2011rz} that reconstructs bulk fields from boundary fields. 

It might be a little surprising that such a smearing function exists, since spatial infinity appears to be causally disconnected from points in the bulk. However, as we have emphasized above, $\spatinf$ refers to the asymptotic dS$_3$ slice that interpolates between $\scrim$ and $\scrip$. By an application of the timelike tube theorem \cite{borchers1961vollstandigkeit,araki1963generalization,Strohmaier:2023opz} we see that, even in a quantum field theory,  it should be possible to reconstruct operators on the entire Cauchy slice at $t = 0$ provided we are given operators on all of $\spatinf$.  In the analysis below, this will be reflected in the fact that the smearing functions are never compactly supported in $\tau$, although they die off doubly exponentially in the limit $\tau \rightarrow \pm \infty$.

We are looking for a function $\smrf$ with the property that when it is smeared with a boundary operator, one obtains a bulk field operator,
\be
\phi(x) = \int \smrf(x,  X') \opspat(X') \dds'.
\ee
In the equation above $x$ is a four-dimensional bulk coordinate that is outside the origin's light cone,  whereas $X'$ is a boundary coordinate on $\spatinf$. 

By matching \eqref{bulkbdry} with \eqref{bdrybdry} we see that the smearing function must satisfy 
\be
\begin{split}
&\int \left(\smrf(x, \tau' + i {\pi \over 2}, \Omega') + (-1)^{\ell} \smrf(x,  -\tau' + i {\pi \over 2}, \Omega') \right)  Y_{\ell}^*(\Omega') d \Omega' \\
&= -m i^{-\ell} j_{\ell}(m r \sinh \tau') e^{i m \cosh \tau' t} Y_{\ell}^*(\Omega).
\end{split}
\ee
We find that this is satisfied by
\be
\smrf(x, \tau', \Omega') = {m \over \pi} \sum_{\ell} k_{\ell}(m r \cosh \tau') e^{m \sinh \tau' t} Y_{\ell}(\Omega') Y_{\ell}^*(\Omega),
\ee
where $k_{\ell}$ is the spherical modified Bessel function \cite[\href{https://dlmf.nist.gov/10.47}{(10.47)}]{NIST:DLMF}.

We note that the $\ell = 0$ term matches the answer found in \cite{Laddha:2022nmj}. We emphasize that the formula above is applicable only in the free theory. It is of interest to generalize this formula to the interacting case, as has been done in AdS \cite{Kabat:2012av}.

\section{A first look at interacting correlators  \label{secbareinteracting}}
In this section, we turn to interacting quantum field theories. We show that, in the presence of interactions, simply smearing correlators with appropriate analytic functions is insufficient to define interacting correlators at $\spatinf$. Contrary to the hope expressed in v1 of \cite{Laddha:2022nmj}, even smeared correlators fall off too slowly to permit extrapolation to $\spatinf$. 

The underlying reason for this is that the large-$\rho$ limit does not automatically put the external fields ``on shell''. Correlation functions at large $\rho$ continue to receive contributions from momenta that might not satisfy $k^2 +m^2 = 0$ and might even be spacelike. The analysis of \ref{subsecextrapplane} tells us that such contributions do not necessarily die off as $e^{-m \rho}$. On the other hand, since these ill-behaved terms come from off-shell contributions, they are not invariant under field redefinitions. 

This suggests a solution to our problem that we will describe in the next section. The bad behaviour at large $\rho$ can be cured through an LSZ-like prescription where one extracts the on-shell part of the correlator before taking the limit to $\spatinf$. This leads to a good extrapolated algebra at $\spatinf$. 

\subsection{Wightman functions and field normalization \label{defwight}}
Consider a local operator $\phi(x)$ that is normalized to have the right matrix elements between one-particle states and the vacuum  \cite{Coleman:2018mew}.
\be
\label{onepartnormalization}
\langle k | \phi(x) |\Omega \rangle = e^{-i k \cdot x}.
\ee
For loop-level computations, we assume that a renormalization scheme has been chosen so that \eqref{onepartnormalization} holds. This convention helps avoid any wave-function renormalization factors in the formulas below.

In the rest of this paper, we study Wightman functions corresponding to matrix elements of products of field operator between two arbitrary states,  $|\Psi_1 \rangle$ and $|\Psi_2 \rangle$ 
\be
\label{genericwightman}
W^{\Psi_1, \Psi_2}(x_1,\ldots,x_n) = \langle \Psi_1 | \phi(x_1),\ldots,\phi(x_n) | \Psi_2 \rangle.
\ee
We denote the Fourier transform of this Wightman function by  $W^{\Psi_1, \Psi_2}(k_1, \ldots, k_n)$. We will also consider the partial Fourier transform that we will denote using 
\[
W^{\Psi_1, \Psi_2}(x_1, \ldots, x_m, k_{m+1}, \ldots, k_n).
\]
When we omit the external states, this means that the expectation value is taken in the vacuum.
\be
\label{vacuumwightman}
W(x_1,\ldots,x_n) = \langle \Omega | \phi(x_1) \ldots \phi(x_n) | \Omega \rangle.
\ee
Here, the vacuum refers to the full interacting vacuum and not the Fock vacuum. 

\subsection{Why Wightman functions?}

In \eqref{genericwightman}, the ordering of operators has been chosen and does not depend on the value of the time coordinate.  The reason for studying Wightman correlators, as opposed to time-ordered correlators is as follows. As explained, operators at $\spatinf$ are obtained by smearing bulk operators with analytic functions before extrapolating them.  However, the presence of theta functions in the time-ordered product spoils the analytic properties of the smearing functions.  
\be
\begin{split}
&\int {\dds}_{1} {\dds}_2 T\{\phi(\rho, X_1) \phi(\rho, X_2)\} \smone(X_1) \smtwo(X_2) \\
&= \int {\dds}_{1} {\dds}_2 \left(\phi(\rho, X_1) \phi(\rho, X_2) \theta(X_1^0 - X_2^0) + \theta(X_2^0 - X_1^0) \phi(\rho, X_2) \phi(\rho, X_1) \right) \smone(X_1) \smtwo(X_2).
\end{split}
\ee
So time-ordering is akin to smearing the product of two bulk operators with a smearing function of two variables that contains a $\theta(X_2^0 - X_1^0)$. Such a smearing function is not analytic either in $X_1$ or in $X_2$. This means that time-ordered correlators cannot be extrapolated to $\spatinf$ even in the free theory.

While time-ordered correlators lead to a simple perturbative expansion, Wightman correlators do not ``miss'' any information. Given  all Wightman functions, one can recover all time-ordered correlators. Moreover, just like time-ordered correlators,  Wightman correlators can also be obtained from the Euclidean theory. This is done by rotating the Euclidean time coordinate for each insertion to Lorentzian time, individually, in the order in which one wants the final Lorentzian operators. In contrast, time-ordered correlators are obtained by rotating all time coordinates to the Lorentzian domain simultaneously \cite{Haag:1992hx}.

\subsection{Large $\rho$ limit}
Starting with the bulk Wightman function \eqref{genericwightman}, consider taking the n\textsuperscript{th} point to $\spatinf$. We smear the point over dS$_3$ with an analytic smearing function, which leads us to study the large $\rho$ limit of
\be
\label{wightintegrated}
W^{\Psi_1, \Psi_2}(x_1,\ldots,x_{n-1}, \rho, \smone) =  \int {\dds}_n W^{\Psi_1, \Psi_2}(x_1,\ldots,x_{n-1}, \rho, X_n) \smone(X_n).
\ee
It is convenient to partially Fourier transform the Wightman function to take this limit
\be
W^{\Psi_1, \Psi_2}(x_1,\ldots,x_{n-1}, \rho, \smone) = \int {\dds}_n  W^{\Psi_1, \Psi_2}(x_1,\ldots,x_{n-1}, k_n) e^{i k_n \cdot \rho X_n} {d^4 k_n \over (2 \pi)^4} \smone(X_n).
\ee
We inserting an identity into the integral written in the form
\be
1 = \int_{-\infty}^{\infty}  {d a \over 2 \pi} \, 2 \pi \delta(k_n^2 + a).
\ee
After writing the measure as
\be
{d^4 k_n \over (2 \pi)^4} = {d \omega_n \over (2 \pi)} |\vec{k}_n|^2 {d |\vec{k}_n| \over 2 \pi} {d^2 \hat{k}_n \over (2 \pi)^2},
\ee
we can do the integral over $|\vec{k}_n|$ to obtain,
\be
\label{wightseparatedtimespace}
W^{\Psi_1, \Psi_2}(x_1,\ldots,x_{n-1}, \rho, \smone) = \int {d^2 \hat{k}_n  \over (2 \pi)^2} {d a \over 2 \pi}  {d \omega_n \over 2 \pi} \theta(\omega_n^2 - a) {\sqrt{\omega_n^2 - a} \over 2}  W^{\Psi_1, \Psi_2}(x_1,\ldots,x_{n-1}, k) I(a,k),
\ee
where 
\be
I(a,k) \equiv \int e^{i k \cdot \rho X_n} \smone(X_n) {\dds}_n,
\ee
and the four-vector, $k$ satisfies $k^2 + a = 0$ and is given by
\be
k = \big(\omega_n, \sqrt{\omega_n^2 - a} \hat{k}_n\big).
\ee

First, consider the case where $a > 0$, which already suffices to establish our result.  The integral reduces to one that we have already studied in section \ref{sectwoptfree}.  Repeating the analysis of section \ref{sectwoptfree}, we find that for $a > 0$, 
\be
I(a,k) \underset{\rho \rightarrow \infty}{\longrightarrow}  -4 \pi \sqrt{\pi \over 2} {1 \over a^{3 \over 4} \rho^{3 \over 2}} e^{-\sqrt{a} \rho}  \times \begin{cases} \sum_{\ell} (\smone_{\ell}(\tau_0 - i {\pi \over 2}) + (-1)^{\ell} \smone_{\ell}(-\tau_0 - i {\pi \over 2}) ) Y_{\ell}^*(\hat{k})& \omega_n > 0 \\
 \sum_{\ell} (\smone_{\ell}(\tau_0 + i {\pi \over 2}) + (-1)^{\ell} \smone_{\ell}(-\tau_0 + i {\pi \over 2}) ) Y_{\ell}^*(-\hat{k}) & \omega_n < 0 \end{cases}
\ee
with $\sqrt{a} \cosh \tau_0 = |\omega_n|$.

In general, we expect the interacting Wightman function to have support for all values of timelike momenta including values of  $a < m^2$. We do not see any reason that these contributions should cancel for generic Wightman correlators. Therefore this analysis shows that the falloff at large $\rho$ is slower than the falloff in the free theory. 

Repeating the analysis of section \ref{sectwoptfree} for spacelike momenta with $k^2 - a = 0$, we find that the $\tau$ integral has saddle points on the real axis. These saddles yield an oscillatory factor of the form $e^{\pm i \sqrt{|a|} \rho}$ at large $\rho$. Therefore the contribution of spacelike momenta to the asymptotic behaviour at large $\rho$ can only be determined once the full dependence of \eqref{wightseparatedtimespace} on $a$ is known. We omit the details here since the contribution of spacelike momenta does not change the conclusion that some terms in the interacting Wightman function die off slower than in the free Wightman function.

We conclude this brief section with a few observations.
\begin{enumerate}
\item
In the free theory, these slow-decaying contributions vanish. This is  because 
all correlators factorize into two-point functions whose Fourier transform is concentrated on mass-shell i.e. on momenta that satisfy $k^2 + m^2 = 0$ in Fourier space.
\item
In the interacting theory, the slow-decaying terms appear at low-orders in perturbation theory. Even if one studies a contact diagram, the structure of perturbation theory (reviewed in Appendix \ref{reviewfeynman}) tells us that external points must be connected to the vertex with four kinds of propagators: time-ordered, anti-time-ordered, Wightman and anti-Wightman. 
However, time-ordered and anti-time-ordered correlators have support {\em everywhere} in momentum space. As a result, the Fourier transform of the interacting Wightman function is not confined to the mass shell.
\item
The slow decaying terms above are not directly related to an S-matrix. Therefore they change under field redefinitions. In the next section, we will show that a generalization of the LSZ-prescription to Wightman functions removes these terms.
\end{enumerate}

\section{An interacting algebra at $\spatinf$ \label{seconshellinteract}}

In the previous section we argued that the Wightman function does not have the correct falloff because its Fourier transform is not concentrated on the mass-shell. In this section, we will make the following proposal. 
\begin{enumerate}
\item
By isolating the on-shell part of the Wightman correlator, through an LSZ-like prescription, one obtains a function that has the right falloff as one approaches $\spatinf$.  We propose that the extrapolated limit of this function should be used to define the algebra at $\spatinf$. 
\item
The algebra at $\spatinf$ is universal in the sense that it is independent of local bulk field redefinitions in the Wightman function --- a property that it shares with the S matrix.
\end{enumerate}
The definition of Wightman functions and the normalization of field operators is the same as the one outlined in section \ref{defwight}. In the next section we will relate correlators at $\spatinf$ to the standard algebra at $i^{\pm}$, and thereby to the S-matrix. 

\subsection{\bf Proposal.}

We now specify what we mean by the ``on-shell'' part of the Wightman function and then explain how it can be extrapolated to $\spatinf$. 

\paragraph{\bf On-shell part of the Wightman function.}
We write the Fourier transformed Wightman function as
\be
\label{wightmanonshell}
W^{\Psi_1, \Psi_2}(k_1,\ldots,k_n) = W^{\Psi_1, \Psi_2}_{\spatinf}(k_1,\ldots,k_n)  + \ldots,
\ee
where $W^{\Psi_1, \Psi_2}_{\spatinf}$, which we will call the ``on-shell Wightman function'', has the form
\be
\label{wightspatinf}
W^{\Psi_1, \Psi_2}_{\spatinf}(k_1,\ldots,k_n) = G^{\Psi_1, \Psi_2}(k_1,\ldots,k_n) (2 \pi) \delta(k_1^2 + m^2) \ldots (2 \pi) \delta(k_n^2 + m^2),
\ee
where $m$ is the physical mass of the particle. The $\ldots$ in \eqref{wightmanonshell} correspond to the remaining terms in the Wightman function whose Fourier transform is not concentrated on the mass-shell. Said in words, the on-shell Wightman function is the part of the Wightman function that has a mass-shell delta function for each external momentum. Note that the delta functions are included in the definition of $W^{\Psi_1, \Psi_2}_{\spatinf}$. 

This on-shell Wightman function can be Fourier transformed back to position space,
\be
W^{\Psi_1, \Psi_2}_{\spatinf}(x_1,\ldots,x_n) = \int W^{\Psi_1, \Psi_2}_{\spatinf}(k_1,\ldots,k_n) e^{i k_1 \cdot x_1 + \ldots i k_n \cdot x_n} \prod_{i} {d^{4} k_i \over (2 \pi)^4}.
\ee
Our proposal is that the algebra at $\spatinf$ should be defined by extrapolating  $W^{\Psi_1, \Psi_2}_{\spatinf}(x_,\ldots, x_n)$, to large $\rho$ in each of its coordinates and smearing with appropriate smearing functions.

An equivalent definition of this on-shell Wightman function is as follows.
\be
\label{altwightinfty}
W^{\Psi_1, \Psi_2}_{\spatinf}(x_1 \ldots x_n) =  \lim_{\epsilon \rightarrow 0}  \int   W^{\Psi_1, \Psi_2}(k_1 \ldots k_n) e^{i k_1 \cdot x_1 + \ldots i k_n \cdot x_n} \prod_i \theta(k_i^2 + m^2 + \epsilon) \theta(\epsilon - k_i^2 - m^2) {d^{4} k_i \over (2 \pi)^4}.
\ee
This integrates the Fourier transform of the Wightman function in a thin shell about $k_i^2 = m^2$ for each momentum. In the limit where the thickness of the shell vanishes, this integral only picks up the delta function in the Fourier transform. 

The Wightman function is expected to have poles on the mass shell. After accounting for the $i \epsilon$ prescription, each such pole can be written as the sum of a principal value and a delta function,  and it is the delta function that contributes to the integral above. However, Wightman functions can contain delta functions without associated poles and these also contribute to the integral. We explain this in more detail, near equation \eqref{decomppole} below.

\begin{figure}[!h]
\begin{center}
\includegraphics[width=0.4\textwidth]{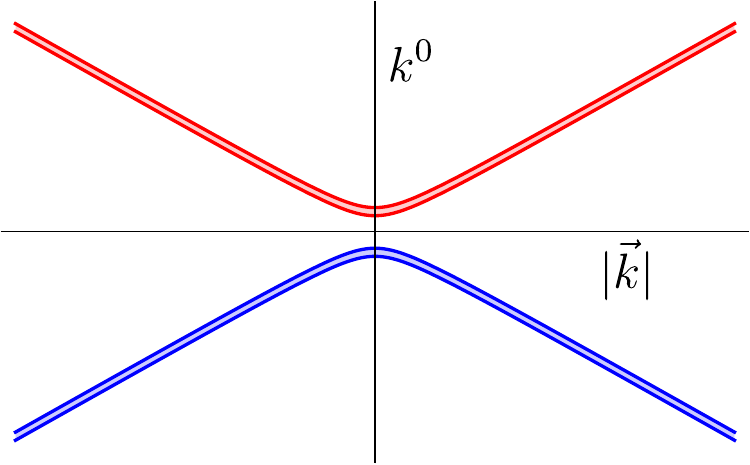}
\caption{$\spatinfann(\vec{k})$ and $\spatinfcre(-\vec{k})$ are obtained by integrating the full Heisenberg operator $\phi(k^0, \vec{k})$ in an infinitesimal range of positive $k^0$ (red region) and negative $k^0$ (blue region) respectively. \label{massshellextraction}}
\end{center}
\end{figure}
The extraction of the on-shell part can be done at the level of a single operator rather than the entire Wightman function. If $\phi(k)$ is the Fourier transform of the full Heisenberg operator, we define
\be
\label{phipmdef}
\begin{split}
&\spatinfann(\vec{k}) = \lim_{\epsilon \rightarrow 0} \int_{\omega_{k}-\epsilon}^{\omega_{k}+\epsilon} \phi(k) {(2 |k^0|) d k^0 \over 2 \pi} ; \\
&\spatinfcre(-\vec{k}) = \lim_{\epsilon \rightarrow 0} \int_{-\omega_{k}-\epsilon}^{-\omega_{k}+\epsilon} \phi(k) {(2 |k^0|) d k^0 \over 2 \pi},
\end{split}
\ee
where $\omega_{k} \equiv \sqrt{\vec{k}^2 + m^2}$. 
In the limit as $\epsilon \rightarrow 0$ only the delta function term in the Fourier transform contributes to the integral. The extraction of these modes is depicted graphically in Fig \ref{massshellextraction}. Note that $\spatinfop(\vec{k})$ are functions of 3-vectors and not 4-vectors. 

Now, we define the operator
\be
\label{phispatinfx}
\phi_{\spatinf}(x) \equiv \int {d^3 \vec{k} \over (2 \pi)^3 2 \omega_{k}} \left( \spatinfann(\vec{k}) e^{i k \cdot x} + \spatinfcre(\vec{k}) e^{-i k \cdot x} \right),
\ee
where the four-vector $k \equiv (\omega_{k}, \vec{k})$. 
Then we can write
\be
W^{\Psi_1, \Psi_2}_{\spatinf}(x_1 \ldots x_n)= \langle \Psi_1| \phi_{\spatinf}(x_1) \ldots \phi_{\spatinf}(x_n) | \Psi_2 \rangle.
\ee

In the free theory, $\phi_{\spatinf}(x)$ coincides with the field itself.  In the interacting theory, the operator $\phi_{\spatinf}(x)$ provides the answer to a natural question: what part of the Heisenberg operator remains concentrated on the mass shell? This is a natural question that could have been asked at the advent of quantum field theory. However, we are not aware of any existing investigations in the literature. In the interacting theory, we will see that $\phi_{\spatinf}(x)$ has an interesting relation to the \inim and \outip operators that are defined at past and future infinity.

\paragraph{\bf Extrapolate limit.}
After smearing with appropriate smearing functions, the analysis of section \ref{subsecextrapplane} tells us that on-shell Wightman functions fall off correctly as one takes $\rho \rightarrow \infty$. The extrapolated on-shell Wightman functions yield correlators at the blowup of spatial infinity
\be
\label{onshellextrap}
\begin{split}
&\langle \Psi_1 | \opspat(\smone_1) \ldots \opspat(\smone_n) | \Psi_2 \rangle \\
&= \lim_{\rho \rightarrow \infty} D(\rho)^n \int {\dds}_1 \ldots {\dds}_n W^{\Psi_1, \Psi_2}_{\spatinf}(\rho, X_1, \ldots, \rho, X_n) \smone_1(X_1) \ldots \smone_n(X_n).
\end{split}
\ee
It is not necessary to take the different $\rho$ coordinates to infinity simultaneously, and we have done so only to simplify the notation. 

At  an operator level we can write
\be
\label{opspatphipm}
\opspat(\smone) =  \int {d^3 \vec{k} \over (2 \pi)^3 2 \omega_{k}} \left(\spatinfann(\vec{k}) \widetilde{\smone}^{+}(\vec{k})  + \spatinfcre(\vec{k}) \widetilde{\smone}^{-}(\vec{k}) \right),
\ee
where $\widetilde{\smone}^{\pm}$ we calculated in subsection \ref{subsecextrapplane}.
The $n$-point correlator on the boundary, displayed on the left hand side of \eqref{onshellextrap}, is obtained by taking the expectation value of the product of such operators between two states.

\paragraph{\bf Comparison with LSZ}
The prescription above isolates the coefficient of the delta function in the Wightman function to obtain a correlator at $\spatinf$.  This is similar in spirit to the LSZ formula that tells us that the S matrix is the residue of a particular pole in the time-ordered correlation function. In time-ordered correlators, delta functions come together with poles since
\be
\label{decomppole}
{1 \over k^2 + m^2 - i \epsilon} = {\rm P.V.}\left({1 \over k^2 + m^2}\right) + i \pi \delta(k^2 + m^2),
\ee
where ${\rm P.V.}$ denotes the principal value. For such propagators, the coefficient of the delta function is proportional to the residue of the pole.

However, Wightman functions can also contain other kinds of external propagators. For instance, the anti-time-ordered propagator has the opposite coefficient of proportionality between the residue and the delta function
\be
\label{decomppoleat}
{1 \over k^2 + m^2 + i \epsilon} = {\rm P.V.}\left({1 \over k^2 + m^2}\right) - i \pi \delta(k^2 + m^2).
\ee
Moreover, when the external propagators are Wightman functions or anti-Wightman functions, delta functions might arise without a corresponding principal-value term.  

The prescription above instructs us to keep track of delta functions from all sources.

\subsection{Perturbative computations}
It is straightforward to write down Feynman rules for Wightman functions. (See Appendix \ref{reviewfeynman}.) These rules are different from the rules for standard time-ordered correlators and instruct us to use a contour with multiple folds. A simple modification of these rules yields the on-shell Wightman function.

The Feynman rules for Wightman correlators make reference to the time-ordered, anti-time-ordered, Wightman and anti-Wightman propagators.  After applying the usual renormalization conditions on the two-point function and, in a scheme, where the wave-function renormalization factors are unity, these propagators have the following form in momentum space:
\be
\label{renormalizedprop}
\begin{split}
&T(k) = {-i \over k^2 + m^2 - i \epsilon} + t(k^2); \qquad \overline{T}(k) = {i \over k^2 + m^2 + i \epsilon} + \bar{t}(k^2); \\
&W(k) = 2 \pi \theta(k^0) \delta(k^2 + m^2) + w(k^2) ; \qquad \overline{W}(k) = 2 \pi \theta(-k^0) \delta(k^2 + m^2) + \bar{w}(k^2),
\end{split}
\ee
where $t,\bar{t}, w, \bar{w}$ are all functions that are regular near $k^2 = m^2$ and $m$ is the physical mass. 

Each of these propagators contains a part proportional to $\delta(k^2 + m^2)$. In the time-ordered and anti-time-ordered propagators, this term arises from the $i \epsilon$ in the denominator. In the case of the Wightman and anti-Wightman function, it appears without a corresponding pole. Moreover, we do not expect a $\delta(k^2 + m^2)$ pole from any other term in perturbation theory except for an external propagator. Therefore,  we see that the on-shell part of the Wightman function is obtained by making the following replacements to {\em external propagators},
\be\label{onshellrules}
\begin{split}
&T(k) \rightarrow \TI(k)= \pi \delta(k^2 + m^2) ; \qquad \overline{T}(k) \rightarrow \ATI(k)= \pi \delta(k^2 + m^2); \\
&W(k) \rightarrow \w(k)= 2 \pi \theta(k^0) \delta(k^2 + m^2); \qquad \overline{W}(k) \rightarrow \wb(k)= 2 \pi \theta(-k^0) \delta(k^2 + m^2).
\end{split}
\ee
All other Feynman rules for Wightman functions, including rules for internal propagators, remain unchanged.

Perturbation theory is designed to compute the Fourier transform of the full Heisenberg operator $\phi(k)$. So, in perturbative computations, we will naturally encounter the four-dimensional Fourier transform of $\phi_{\spatinf}(x)$  that is a combination of the on-shell operators defined in \eqref{phipmdef}.
\be
\label{phispatinfphipm}
\phi_{\spatinf}(k) \equiv \left(\spatinfann(\vec{k}) \theta(k^0)  + \spatinfcre(-\vec{k}) \theta(-k^0) \right) (2 \pi) \delta(k^2 + m^2).
\ee
Note that $\phi_{\spatinf}(k)$ depends on a four-vector although its support is concentrated on the mass shell. 

\subsection{Two-point function}
Our procedure for extracting an on-shell correlator is also valid for the vacuum two-point functions without any modification. This case might seem confusing since the two-point function is conventionally written with only one mass-shell delta function. However, the overall energy-momentum conserving delta function automatically puts the other momentum on shell.

The two-point vacuum Wightman correlator is
\be
\begin{split}
W(k_1, k_2) &= \int \langle \Omega| \phi(x_1) \phi(x_2) | \Omega \rangle e^{-i k_1 \cdot x_1 - i k_2 \cdot x_2} d^4 x_1 d^4 x_2 \\ &=  \left[ 2 \pi \theta(k_1^0) \delta(k_1^2 + m^2)  + w(k_1^2) \right]  (2 \pi)^4 \delta^4(k_1 + k_2),
\end{split}
\ee
where we remind the reader that $w$, introduced previously in \eqref{renormalizedprop}, is some function that is regular at $k_1^2 + m^2 = 0$.
From here, we can read off 
\be
W_{\spatinf}(k_1, k_2) = 2 \pi  (2 k_1^0) \theta(k_1^0) \theta(-k_2^0) \delta(k_1^2 + m^2) 2 \pi \delta(k_2^2 + m^2) (2 \pi)^3 \delta^3(\vec{k}_1 + \vec{k}_2),
\ee
where we have rewritten one of the energy-momentum delta functions as a mass-shell delta function and discarded the regular term. We remind the reader that the on-shell Wightman function is {\em not} just the coefficient of the mass-shell delta functions --- it includes the delta functions themselves. Note that the two-point on-shell Wightman function coincides with the free Wightman function even in the presence of interactions.

It is also useful to consider the case where one of the external states is a single-particle state; the other external state is the vacuum; and we have one insertion of the field at spatial infinity. Then we see, from the normalization \eqref{onepartnormalization} that
\be
\label{onepartmatrixelem}
\begin{split}
&\langle k | \spatinfcre(\vec{k}') | \Omega \rangle = (2 \pi)^3 2 \omega_{k} \delta^3(\vec{k} - \vec{k}'); \qquad \langle \Omega | \spatinfcre(\vec{k}') | k  \rangle = 0 \\
&\langle \Omega | \spatinfann(\vec{k}') | k \rangle = (2 \pi)^3  2 \omega_{k} \delta^3(\vec{k} - \vec{k}'); \qquad \langle k | \spatinfann(\vec{k}') | \Omega \rangle = 0. \end{split} 
\ee

\section{Relation between operators at $\spatinf$ and $i^{\pm}$ \label{secspatinfinout}}
In this section, we relate the algebra of operators at $\spatinf$ to the algebra of \inim and \outip operators defined at $i^{-}$ and $i^{+}$ respectively. Remarkably, it will turn out that operators at $\spatinf$ defined via our prescription are naturally related to the arithmetic mean of \inim and \outip operators.

Let $a_{\vec{k}}, a_{\vec{k}}^{\dagger}$ be the \inim annihilation and creation operators. This means that, in the far past --- more precisely, as one approaches $i^{-}$ \cite{Campiglia:2015kxa} ---  the field can be expanded linearly in these operators. Let $b_{\vec{k}}, b_{\vec{k}}^{\dagger}$ be the \outip operators that play the same role in the far future, as one approaches $i^+$. 

We will show that
\be
\label{momopspatinout}
\momopspat(k) = (2 \pi) \delta(k^2 + m^2) \left[  \theta(k^0) {1 \over 2} (a_{\vec{k}} + b_{\vec{k}}) +  \theta(-k^0) {1 \over 2} (a_{-\vec{k}} + b_{-\vec{k}})^{\dagger}  \right].
\ee
This means that
\be
\label{spatinfinout}
\spatinfann(\vec{k}) = {1 \over 2} \left(a_{\vec{k}} + b_{\vec{k}} \right); \qquad \spatinfcre(\vec{k})= {1 \over 2} \left(a_{k}^{\dagger} + b_{\vec{k}}^{\dagger}\right).
\ee
Our proof will be diagrammatic.  Consider an arbitrary matrix element of the following form.
\be
\label{matrixelem}
M = \lout \vec{q}_1,\ldots,\vec{q}_n | \momopspat(k) | \vec{p}_1,\ldots,\vec{p}_m \rin.
\ee
The advantage of considering such a matrix element is that it can be computed using a {\em single-fold contour.}  Therefore the diagrams that contribute to this matrix element reduce to ordinary Feynman diagrams, familiar from time-ordered perturbation theory. This is not difficult to understand: by definition, the states on the extreme right are created by fields in the far past; the states on the left are created by fields in the far future; the prescription \eqref{wightspatinf} instructs us to compute the matrix elements of $\phi(k)$ and pick out the term that contains $\delta(k^2 + m^2)$. 

The basis of matrix elements displayed in \eqref{matrixelem} is complete. Therefore it is sufficient to show that formula \eqref{momopspatinout} holds for these matrix elements. 

\paragraph{\bf Connected diagrams}
First let us consider diagrams that contribute to \eqref{matrixelem} where  $\momopspat(k)$ connects to an internal vertex. 
\begin{figure}[!h]
\begin{center}
       \includegraphics[height=0.3\textheight]{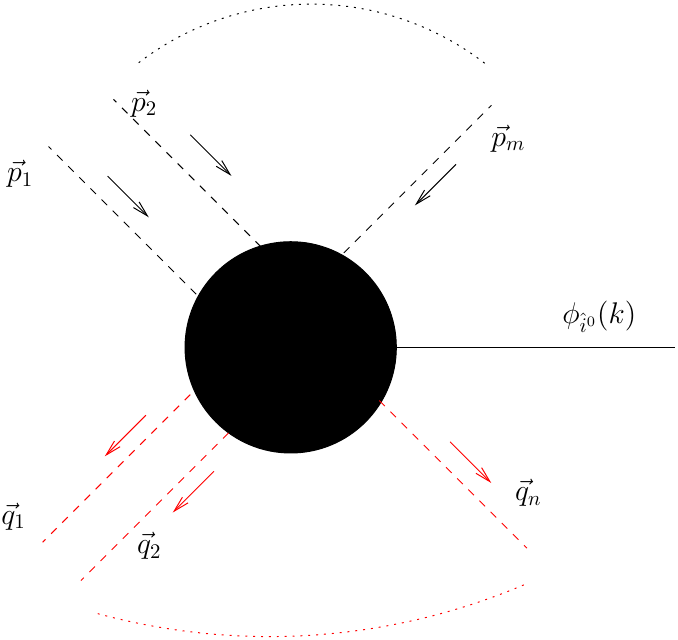}
        \caption{A diagram where $\momopspat(k)$ connects to an internal vertex in the blob, which then connects to the \inim and \outip legs. \label{connected}}
\end{center}
\end{figure}
In such a diagram the insertion of $\momopspat(k)$ is almost precisely the same as the insertion of an external line in the matrix element, except that the propagator that connects the insertion to the internal vertex carries an 
additional factor of $\pi \delta(k^2 + m^2)$. In the case where $k^0 > 0$, the insertion behaves like an additional \outip state, whereas when $k^0 < 0$, the insertion behaves like an additional \inim state. However, in the latter case, the momentum of the \inim state is $-\vec{k}$. We can write this contribution as
\be
\label{momopspatcon}
\begin{split}
M_{\rm con} = {\pi \delta(k^2 + m^2) } \Big[&\theta(k^0) \lout \vec{q}_1,\ldots,\vec{q}_n \vec{k} | \vec{p}_1, \ldots, \vec{p}_m \rinp \\ + &\theta(-k^0)  \lout \vec{q}_1,\ldots,\vec{q}_n |  -\vec{k}, \vec{p}_1, \ldots, \vec{p}_m \rinp \Big].
\end{split}
\ee
The $'$ on the correlator means that we have excluded diagrams where the insertion of $\momopspat$ contracts directly with one of the \inim or \outip states. This means
\be
\begin{split}
 \lout \vec{q}_1,\ldots,\vec{q}_n | -\vec{k}, \vec{p}_1,\ldots,\vec{p}_m \rinp =  \lout\vec{q}_1,\ldots,\vec{q}_n | -\vec{k}, \vec{p}_1,\ldots,\vec{p}_m  \rin& \\
-  2 \omega_{k} \sum_i (2 \pi)^3 \delta^{3}(\vec{k} + \vec{q}_i) \lout \vec{q}_1,\ldots,\cancel{\vec{q}_i},\ldots,\vec{q}_n | \vec{p}_1,\ldots,\vec{p}_m \rin&,
\end{split}
\ee
and likewise,
\be
\begin{split}
 \lout \vec{q}_1,\ldots,\vec{q}_n, \vec{k} |  \vec{p}_1,\ldots,\vec{p}_m \rinp =   \lout \vec{q}_1,\ldots,\vec{q}_n, \vec{k}| \vec{p}_1,\ldots,\vec{p}_m \rin & \\ - 2 \omega_{k} \sum_i (2 \pi)^3 \delta^{3}(\vec{k}-\vec{p}_i)  \lout \vec{q}_1,\ldots,\vec{q}_n | \vec{p}_1,\ldots,\cancel{\vec{p}_i},\ldots,\vec{p}_m \rin&.
\end{split}
\ee
When $\vec{k}$ does not coincide with one of the $\vec{q}_i$ or the $\vec{p}_i$, we have
\be
M = M_{\rm con}, \qquad \vec{k} \notin \{\vec{q}_i \}, \vec{k} \notin \{\vec{p}_j\}.
\ee
The insertion of the right hand side of \eqref{momopspatinout} manifestly yields the same result as $M_{\rm con}$ for this case. 

\begin{figure}[t!]
    \centering
~     \begin{subfigure}[t]{0.48\textwidth}
        \centering
        \includegraphics[height=3in]{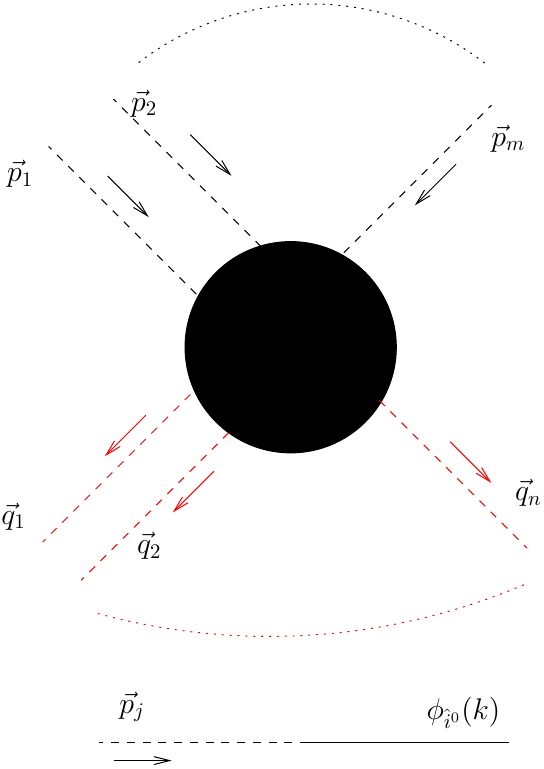}
        \caption{A diagram where $\momopspat(k)$ is contracted directly with an \inim leg.}
    \end{subfigure}
~
        \begin{subfigure}[t]{0.48\textwidth}
        \centering
        \includegraphics[height=3in]{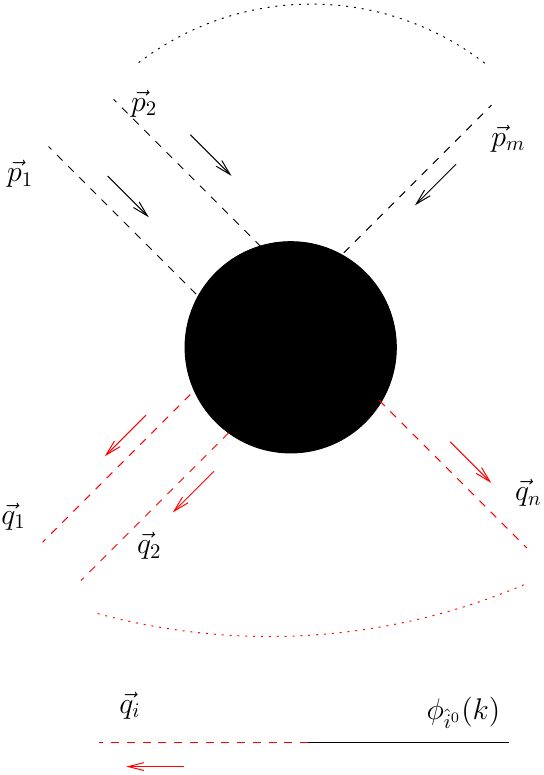}
        \caption{A diagram where $\momopspat(k)$ is contracted directly with an \outip leg.}
    \end{subfigure}
\caption{Two kinds of disconnected diagrams that contribute to $\langle \vec{p}_1,\ldots,\vec{p}_n | \momopspat(k) | \vec{q}_1,\ldots,\vec{q}_m \rangle$ \label{figmomopspatdis}}
\end{figure}

\paragraph{\bf Disconnected diagrams}
To complete the proof of formula \eqref{momopspatinout}, we must also consider the delta-function contributions that arise when the momentum $k$ coincides with one of the incoming or outgoing momenta. These contributions correspond to the disconnected diagrams depicted in Figure \ref{figmomopspatdis}.
We consider two kinds of diagrams: those where $\momopspat(k)$ connects directly to one of the \outip legs, and those where $\momopspat(k)$ connects directly to one of the \inim legs.
Using the normalization in \eqref{onepartmatrixelem}, we see that these diagrams yield a contribution
\be
\label{momopspatdis}
\begin{split}
M_{\rm dis} = (2 \pi)  \delta(k^2 + m^2) &2 \omega_{k} \Big[\sum_{i} \theta(-k^0) (2 \pi)^3 \delta^3(\vec{k} +  \vec{q}_i) \lout \vec{q}_1,\ldots,\cancel{\vec{q}_i},\ldots,\vec{q}_n | \vec{p}_1,\ldots,\vec{p}_m \rin \\ &+ \sum_j \theta(k^0)(2 \pi)^3 \delta^3(\vec{k}- \vec{p}_i) \lout \vec{q}_1,\ldots,\vec{q}_n |  \vec{p}_1,\ldots,\cancel{\vec{p}_j},\ldots,\vec{p}_m \rin  \Big].
\end{split}
\ee

Adding \eqref{momopspatdis} and \eqref{momopspatcon} and noting that the mass-shell delta function in \eqref{momopspatdis} is multiplied with $(2 \pi)$ as opposed to $\pi$ in \eqref{momopspatcon}, we see that
\be
\begin{split}
M = M_{\rm con} + M_{\rm dis} &= {\pi \delta(k^2 + m^2) } \\ \times \Bigg[ &\theta(k^0)\lout \vec{q}_1,\ldots,\vec{q}_n, \vec{k} | \vec{p}_1,\ldots,\vec{p}_m \rin  \\ &+ \theta(-k^0) \lout \vec{q}_1,\ldots,\vec{q}_n | -\vec{k}, \vec{p}_1,\ldots,\vec{p}_m \rin  \\
&+ 2 \omega_{k} \sum_{i} \theta(-k^0) (2 \pi)^3 \delta^3(\vec{k} +  \vec{q}_i) \lout \vec{q}_1,\ldots,\cancel{\vec{q}_i},\ldots,\vec{q}_n | \vec{p}_1,\ldots,\vec{p}_m \rin  \\
&+ 2 \omega_{k} \sum_j \theta(k^0)(2 \pi)^3 \delta^3(\vec{k}- \vec{p}_j) \lout \vec{q}_1,\ldots,\vec{q}_n |  \vec{p}_1,\ldots,\cancel{\vec{p}_j},\ldots,\vec{p}_m \rin \Bigg].
\end{split}
\ee
But this is precisely what one would obtain by inserting the right hand side of \eqref{momopspatinout} in \eqref{matrixelem}. This completes our proof of \eqref{momopspatinout}. 

\subsection{An intermediate discussion}
Asymptotic observables with a general sequence of insertions of the \inim and \outip creation and annihilation operators were studied in \cite{Caron-Huot:2023vxl}.\footnote{Note that the LSZ prescription discussed in \cite{Caron-Huot:2023vxl} differs from ours. \cite{Caron-Huot:2023vxl} studied out-of-time-order correlators of the standard LSZ currents. These are differences of the \inim and \outip operators instead of averages.} The S matrix is a specific observable within this class, where we act with the out-annihilation operators on the left, and the in-creation operators on the right.
\be
\lout \vec{q}_1,\ldots,\vec{q}_n | \vec{p}_1,\ldots,\vec{p}_m \rin = \langle \Omega |  b_{\vec{q}_1} \dots b_{\vec{q}_n} a_{\vec{p}_1}^{\dagger} \dots a_{\vec{p}_m}^{\dagger} | \Omega \rangle.
\ee

However, it is possible to consider correlators with a general string of such operators such as 
\[
 \langle \Omega |  b_{\vec{k}_1} \ldots b_{\vec{k}_i} a_{\vec{k}_{i+1}} \ldots a_{\vec{k}_j} b^{\dagger}_{\vec{k}_{j+1}} \ldots b_{\vec{k}_m}^{\dagger}   a^{\dagger}_{\vec{k}_{m+1}} \ldots a^{\dagger}_{\vec{k}_n} | \Omega \rangle.
\]
The choice of ordering above is made only for purposes of illustration, and many interesting correlators involve the \inim and \outip creation and annihilation operators in all sorts of orderings.

 For most orderings, including the one displayed above, the correlator cannot immediately be converted to a S-matrix because the commutation relations of the \inim and \outip operators are complicated. Recall that
\be
b_{\vec{k}} = S^{\dagger} a_{\vec{k}} S; \qquad b_{\vec{k}}^{\dagger} = S^{\dagger} a_{\vec{k}}^{\dagger} S,
\ee
where $S$ is the S-matrix.

Nevertheless, by inserting a complete set of intermediate states, or by using the relation above, we can always express observables with arbitrary insertions in terms of the S-matrix, its conjugate and powers of these operators.  This is explained in more detail in \cite{Caron-Huot:2023vxl}; in fact, one of the motivations of \cite{Caron-Huot:2023vxl} was to explain that some observables, which are nonlinear in the S matrix, can be represented linearly as correlators of the form above. We will work out some simple examples in the next section. 

We see that the result \eqref{momopspatinout} allows us to express arbitrary correlators at spatial infinity in terms of such strings of operators. This is because correlators of $\opspat(\smone)$ can be related to correlators of $\spatinfop$ using \eqref{opspatphipm}. 
\be
\label{spatinftobdry}
\begin{split}
&\langle \Psi_1 | \opspat(\smone_1) \ldots \opspat(\smone_n) | \Psi_2 \rangle = \int \prod_{i} {d^3 \vec{k}_i \over (2 \pi)^3 2 \omega_{k_i}} {\cal I}; \\
&{\cal I} = \langle \Psi_1 | \left(\spatinfann(\vec{k}_1) \widetilde{\smone}_1^{+}(\vec{k}_1)  + \spatinfcre(\vec{k}_1) \widetilde{\smone}_1^{-}(\vec{k}_1) \right) \ldots \left(\spatinfann(\vec{k}_n) \widetilde{\smone}_n^{+}(\vec{k}_n)  + \spatinfcre(\vec{k}_n) \widetilde{\smone}_n^{-}(\vec{k}_n) \right) | \Psi_2 \rangle.
\end{split}
\ee
The quantum field theory computation here  is simply
\[
\langle \Psi_1 | \spatinfop(\vec{k}_1) \ldots \spatinfop(\vec{k}_n) | \Psi_2 \rangle.
\]
Once we are given this quantity, we merely need to choose an appropriate set of smearing functions and integrate the result against them to obtain the correlator at spatial infinity. 

Since the operators $\spatinfop(\vec{k}_i)$ that appear above are arithmetic means of the \inim and \outip creation operators, it is natural to state that the arithmetic means of \inim and \outip operators provide a natural basis of operators at $\spatinf$.

\section{Sample computations \label{secsample}}
In this section, we present results for several sample computations. These involve correlators of boundary operators in the vacuum and also between non-trivial \inim and \outip states. We will not go through the exercise of explicitly picking smearing functions and integrating our answers for correlators of $\spatinfop(\vec{k})$ against them.  Although this procedure is straightforward in principle,  these integrals often cannot be done analytically.  It might be interesting to perform these integrals for specific simple choices of smearing functions and examine their structure. 

We start with a general result on the reality properties of Wightman functions, which tells us that the sum of all contact diagrams has no on-shell part at tree level. This observation will be useful in understanding the structure of correlators that we obtain. 

We then describe  results for $4$-point, $5$-point and $6$- point correlators in a theory with massive scalar fields.  Each of the correlators can be computed in two different ways. First, one may use  the representation of $\momopspat(k)$ in terms of \inim and \outip fields to relate the correlator to the S-matrix and its powers.  Alternately, one may simply compute the truncated Wightman function using the Feynman rules described in Appendix \ref{reviewfeynman}. We direct the reader to Appendix \ref{details_of_comp} for further details.

A notational point must be kept in mind when one compares the two methods above. In perturbation theory, it is natural to work in terms of $\phi_{\spatinf}(k)$. According to \eqref{phispatinfphipm}, this is a linear combination of $\spatinfann(\vec{k})$ and $\spatinfcre(-\vec{k})$. The sign reversal is standard in perturbation theory: all momenta are marked as outgoing in Feynman diagrams but an \inim momentum marked as $k$ corresponds to a physical four-momentum $-k$.

\subsection{Reality properties of Wightman functions in momentum space}
We start by proving the following simple, but perhaps unfamiliar, statement: the Fourier transform of vacuum-to-vacuum Wightman functions of real scalar fields is real.
\be
\label{realitywight}
\langle \Omega | \phi(k_1) \ldots \phi(k_n) | \Omega \rangle^* = \langle \Omega | \phi(k_1) \ldots \phi(k_n) | \Omega \rangle.
\ee
This property follows from CPT invariance. Under CPT conjugation \cite{streater2016pct} we have
\be
\langle \Omega | \phi(x_1) \ldots \phi(x_n) | \Omega \rangle= \langle \Omega | \phi(-x_n) \ldots \phi(-x_1) | \Omega \rangle,
\ee
or, in momentum space
\be
\langle \Omega | \phi(k_1) \ldots \phi(k_n) | \Omega \rangle= \langle \Omega | \phi(-k_n) \ldots \phi(-k_1) | \Omega \rangle.
\ee
Upon conjugating and noting that $\phi(k)^{\dagger} = \phi(-k)$, we find the result \eqref{realitywight}.

\subsubsection{Vanishing contact diagrams \label{subsecvanish}}
The result above implies that, at tree-level, an interaction of the form $\lambda {\phi^n \over n!}$ never contributes to a $n$-point correlator. One might have expected a nontrivial contribution through a contact diagram but we will now argue that the sum of all such contact diagrams must vanish. 

\begin{figure}[H]
		\centering
		\includegraphics[scale=0.25]{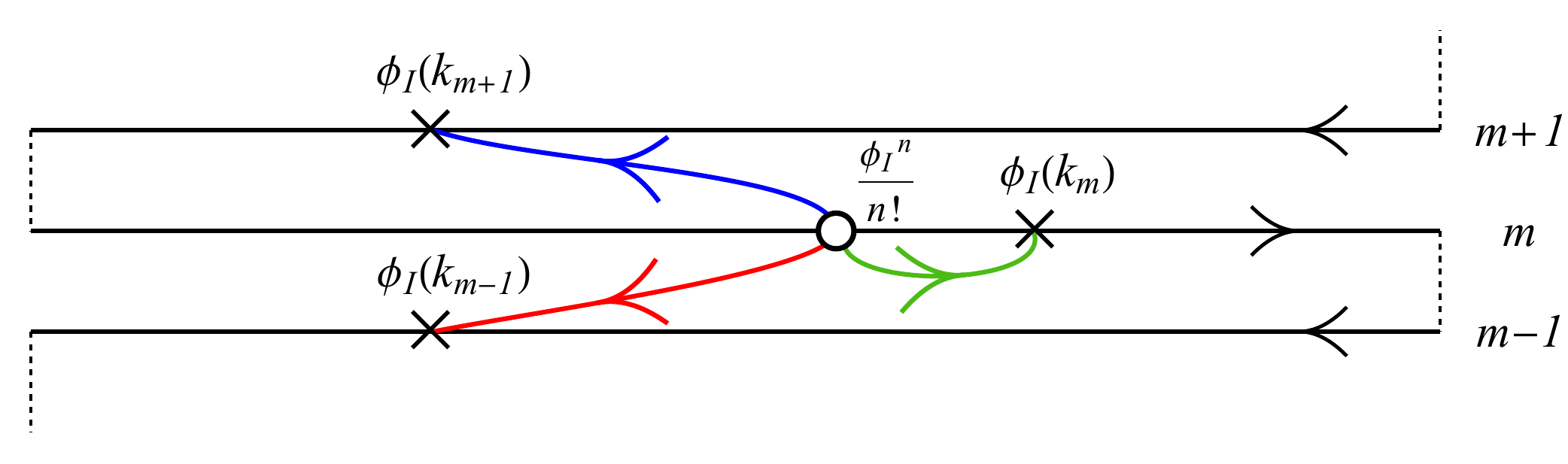}

	\caption{ Contour lines for a contact diagram. There is only one interaction point with an interaction term ${\phi_I^n \over n!}$ that connects with all external points. This interaction point will run through all the contour lines. In this figure we have only shown one particular configuration.}
	\label{ncontact}
\end{figure}
Consider evaluating the contribution of a $n$-point interaction to a $n$-point correlator. Using the rules of Wightman perturbation theory explained in appendix \ref{reviewfeynman}, we consider the contour displayed in Figure \ref{ncontact}. 
The single interaction point can be anywhere on the folds of the contour and the vertex factor associated with the interaction is $(\pm i \lambda)$. The leading sign depends on whether the vertex is on an odd or an even fold of the contour. This interaction point couples to the external points through external propagators. The external propagators can be time-ordered, anti-time-ordered, Wightman or anti-Wightman. However, when we are computing the on-shell part of a Wightman function, we are instructed to use the truncated propagators, $\TI(k), \ATI(k), \w(k), \wb(k)$ displayed in \eqref{onshellrules}. All of these are real. Therefore, after accounting for the vertex factor,  each diagram contributes a purely imaginary term to the correlator. Since the correlator is real, the sum of all diagrams must vanish. 

In perturbation theory, this happens through a nontrivial set of cancellations. Since the vertex can be on different folds, we get terms with differing signs. Each term also has a different combination of external propagators following the Feynman rules explained in Appendix \ref{reviewfeynman}. Nevertheless, the sum of all diagrams always vanishes. The interested reader can look at sections \ref{appsubsecthreept} or sections \ref{appsubsecfourpt} to see the details of these cancellations for three-point and four-point contact interactions.

\subsection{$4$-point vacuum correlator}
We will now derive an expression for the four-point vacuum correlator at spatial infinity.  It is convenient to start by studying insertions of $\spatinfop(\vec{k})$. We will later combine all these possibilities to obtain the expression for smeared operators at $\spatinf$. 

We are required to study the following matrix elements
\be
M_4^{\pm, \pm} = \langle \Omega | \spatinfann(\vec{k}_{1}) \spatinfop(\vec{k}_{2}) \spatinfop(\vec{k}_{3})  \spatinfcre(\vec{k}_{4}) | \Omega \rangle.
\ee
The energy of the vacuum cannot be lowered, and so the correlator vanishes unless the insertion on the extreme left has positive frequency and the insertion on the extreme right has negative frequency. 

Using the expression \eqref{spatinfinout} and noting that one-particle states are unique (i.e. \inim one particle states are the same as \outip one particle states), we see that
\be
M_4^{\pm, \pm} = \langle \vec{k}_{1} | \spatinfop(\vec{k}_2)  \spatinfop(\vec{k}_3) | \vec{k}_{4} \rangle.
\ee
Now we note the relations
\be
\begin{split}
&a_{\vec{k}_{3}}^{\dagger} | \vec{k}_{4} \rangle = |\vec{k}_{3}, \vec{k}_{4} \rin; \qquad b_{\vec{k}_{3}}^{\dagger} | \vec{k}_{4} \rangle = |\vec{k}_{3}, \vec{k}_{4} \rout; \\
&a_{\vec{k}_{3}}| \vec{k}_{4} \rangle = b_{\vec{k}_{3}} | \vec{k}_{4} \rangle = (2 \pi)^3 \delta^3(\vec{k}_3 - \vec{k}_4) (2 \omega_{k_3}) |0 \rangle,
\end{split}
\ee
and their obvious Hermitian conjugates that apply to bras
\be
\begin{split}
&\langle \vec{k}_{1} | a_{\vec{k}_{2}} = \lin \vec{k}_{1}, \vec{k}_{2} |; \qquad  \langle \vec{k}_{1} | b_{\vec{k}_{2}} = \lout \vec{k}_{1}, \vec{k}_{2} |; \\
&\langle \vec{k}_{1} | a_{\vec{k}_{2}}^{\dagger}  = \langle \vec{k}_{1} | b_{\vec{k}_{2}}^{\dagger} =  (2 \pi)^3 \delta^3(\vec{k}_1 - \vec{k}_2) (2 \omega_{k_1}) \langle 0|.
\end{split}
\ee
We also recall our normalization
\be
\begin{split}
D_1 &= \lin \vec{k}_{1}, \vec{k}_{2} | \vec{k}_{3}, \vec{k}_{4} \rin = \lout \vec{k}_{1}, \vec{k}_{2} | \vec{k}_{3}, \vec{k}_{4} \rout \\ &= (4 \omega_{k_{1}} \omega_{k_{2}}) (2 \pi)^6 \big(\delta^3(\vec{k}_1 - \vec{k}_3) \delta^3(\vec{k}_2 - \vec{k}_4)  + \delta^3(\vec{k}_1 - \vec{k}_4) \delta^3(\vec{k}_2 - \vec{k}_3) \big).
\end{split}
\ee
Therefore,
\be
M_4^{+-} = {1 \over 4} {\lin \vec{k}_{1}, \vec{k}_{2} | \vec{k}_{3},\vec{k}_{4} \rout} + {1 \over 4} {\lout \vec{k}_{1}, \vec{k}_{2} | \vec{k}_{3},\vec{k}_{4} \rin}  + {1 \over 2} D_1,
\ee
and 
\be
 M_4^{-+} = D_2 \equiv  (4 \omega_{k_{1}} \omega_{k_{3}}) (2 \pi)^6 \delta^3(\vec{k}_1 - \vec{k}_2) \delta^3(\vec{k}_3 - \vec{k}_4),
\ee
whereas $M_4^{++} = M_4^{--} = 0$. 
We also have
\be \label{comconjugate}
{_\text{out}\langle \vec{k}_{1}, \vec{k}_{2} | \vec{k}_{3},\vec{k}_{4} \rin} = \big({_\text{in}\langle \vec{k}_{3}, \vec{k}_{4} | \vec{k}_{1},\vec{k}_{2} \rout}\big)^*,
\ee
and for real scalar fields, we have from CPT invariance,
\be \label{cpt}
{_\text{in}\langle \vec{k}_{3}, \vec{k}_{4} | \vec{k}_{1},\vec{k}_{2} \rout} = {_\text{in}\langle \vec{k}_{1}, \vec{k}_{2} | \vec{k}_{3},\vec{k}_{4} \rout}.
\ee
So we can write,
\be
\begin{split}
M_4^{+-} &= {1 \over 2}\Re\Big( {\lout \vec{k}_{1}, \vec{k}_{2} | \vec{k}_{3},\vec{k}_{4} \rin}\Big)  + {1 \over 2} D_1\\
&= -{1 \over 2}  \text{Im}\big(T_{\vec{k}_1, \vec{k}_2 \leftarrow \vec{k}_3 \vec{k}_4} \big) + D_1,
\end{split}
\ee
where the $T$ matrix is related to the S-matrix through  $S = 1 + i T$. In the free limit, the $T$ matrix vanishes and we are naturally left with the disconnected contributions.

Given a set of smearing functions, and a particular theory,  we need to compute the $T$-matrix and insert the correlators above into the formula \eqref{phispatinfx} to obtain correlators at $\spatinf$. 
\be
\langle \Omega | \opspat(\smone_1) \opspat(\smone_2) \opspat(\smone_3) \opspat(\smone_4) | \Omega \rangle = {\cal C} + {\cal D},
\ee
where the connected part is
\be
{\cal C} = -\int \widetilde{\smone}_1^{+}(\vec{k}_1)  \widetilde{\smone}_2^{+}(\vec{k}_2)  \widetilde{\smone}_3^{-}(\vec{k}_3)  \widetilde{\smone}_4^{-}(\vec{k}_4)  \big({1 \over 2} \Im T_{\vec{k}_1, \vec{k}_2 \leftarrow \vec{k}_3 \vec{k}_4} \big) \prod_i {d^3 \vec{k}_{i} \over (2 \pi)^3 (2 \omega_{k_i})} , 
\ee
and the disconnected part is
\be
\begin{split}
{\cal D} = \int {d^{3}\vec{k}_1 d^{3} \vec{k}_2 \over (2 \pi)^6 (2 \omega_{k_1} 2 \omega_{k_2})} \Big[&\widetilde{\smone}_1^{+}(\vec{k}_1) \widetilde{\smone}_4^{-}(\vec{k}_1) \widetilde{\smone}_2^{+}(\vec{k}_2) \widetilde{\smone}_3^{-}(\vec{k}_2) + \widetilde{\smone}_1^{+}(\vec{k}_1)\widetilde{\smone}_3^{-}(\vec{k}_1) \widetilde{\smone}_2^{+}(\vec{k}_2)  \widetilde{\smone}_4^{-}(\vec{k}_2)  \\ &+ \widetilde{\smone}_1^{+}(\vec{k}_1)\widetilde{\smone}_2^{-}(\vec{k}_1) \widetilde{\smone}_3^{+}(\vec{k}_2)  \widetilde{\smone}_4^{-}(\vec{k}_2) \Big].
\end{split}
\ee

The formula above is general and agnostic to the specific interactions. We now describe a few special cases.

First consider a simple real scalar with an interaction term  $\frac{\lambda}{3!}\phi^3(x)$. At tree level, one might have obtained an imaginary part for the $T$ matrix at ${\cal O}[\lambda^2]$ through a 4-point exchange diagram when the internal propagator goes on shell. However, this term vanishes since when the external particles are on shell, the internal line cannot be on shell due to energy momentum conservation: we cannot simultaneously satisfy $k_1^2+m^2 = 0$, $k_2^2+m^2 = 0$, and $(k_1+k_2)^2+m^2 = 0$. So the connected part vanishes at $\Or[\lambda^2]$. This conclusion can be checked by directly examining the on-shell part of the four-point Wightman function. (See Appendix \ref{details_of_comp}.) In this calculation some of the cancellations seem rather miraculous although the formula above makes it clear why they should occur. 

Next, consider a self interaction $\frac{g}{4!}\phi^4$. We have already explained that the contact diagram does not provide a nonzero contribution to the correlator at $\spatinf$. However, we can get a  non-zero answer at loop level. This is simply the textbook computation of the imaginary part of the one-loop diagram. However, as above, we can also check it by performing the full perturbative analysis for the Wightman function. A rather involved computation (see Appendix \ref{details_of_comp}) yields
\be
	\begin{split}
			&\langle \Omega | \momopspat(k_1) \momopspat(k_2) \momopspat(k_3)\momopspat(k_4) | \Omega \rangle_{\text{connected}}
		\\ =&
		- \frac{g^2}{64 \pi} (2 \pi)^4 \delta^4\Big(\sum_{i=1}^{4}k_i\Big)	\Big(\prod_{i}(2 \pi)\delta(k_i^2+m^2)\Big)\theta(k_1^0)\theta(k_2^0)\theta(-k_3^0)\theta(-k_4^0) \sqrt{1+\frac{4 m^2}{(k_1+k_2)^2}}.
\end{split}
\ee
This shows, as expected, that it is only $M^{+-}$ matrix element whose connected part is nonzero. Moreover, its value is proportional to the imaginary part of the $T$-matrix. We remind the reader that to compare perturbative correlators of $\momopspat(k)$ with expressions involving the T-matrix, one must flip the sign of $\vec{k}$ when $k^0$ is negative. 

\subsection{$4$-point non-vacuum state correlator }
 We can also study correlators with arbitrary \inim and \outip states and additional insertions of operators at spatial infinity. For instance, consider preparing a two-particle \inim state, $|\vec{p}_1, \vec{p}_2 \rin$  at  $i^-$, inserting an operator at $\spatinf$ and studying the amplitude to obtain the one-particle state $\langle \vec{q} \, |$ at $i^+$.  

In terms of $\spatinfop(\vec{k})$, only one possible choice for the frequency of the insertion yields a nonzero answer. 
\be
\langle \vec{q} \, |  \spatinfann(\vec{k}) | \vec{p}_1,\vec{p}_2 \rin = {1 \over 2} \lout \vec{q}, \vec{k} \,| \vec{p}_1, \vec{p}_2 \rin; \qquad \langle \vec{q} \,| \spatinfcre(\vec{k}) | \vec{p}_1, \vec{p}_2 \rin = 0.
\ee
Given a smearing function, this immediately yields correlators at $\spatinf$ using the formula \eqref{opspatphipm}
\be
\langle \vec{q}| \opspat(\smone) | \vec{p}_1 \vec{p}_2 \rin = \int {d^3 \vec{k} \over (2 \pi)^3 (2 \omega_k) } {1 \over 2}  \lout \vec{q}, \vec{k}| \vec{p}_1, \vec{p}_2 \rin  \widetilde{\smone}^{+}(\vec{k}).
\ee

\subsection{$5$-point vacuum correlator \label{fivepointvac}}
We now turn to the vacuum five-point correlator. In this subsection, we stop our analysis once we have computed the on-shell Wightman function; its conversion into a correlator at $\spatinf$ follows the same pattern as above. Moreover, for simplicity, we only display the connected term. The answers below are complete when the $\vec{k}_i$ are distinct and the disconnected terms can easily be worked out separately. 

In principle, the on-shell five-point Wightman function can be broken up into 32 separate frequency components. However, only two of these components are nonvanishing when the $\vec{k}_i$ are distinct.
\be
	M_5^{\pm} = \langle \Omega | \spatinfann(\vec{k}_{1})\spatinfann(\vec{k}_{2})  \spatinfop(\vec{k}_{3})\spatinfcre(\vec{k}_{4})\spatinfcre(\vec{k}_{5}) | \Omega \rangle.
\ee

We have
\be
\begin{split}
&\spatinfcre(\vec{k}_{4})\spatinfcre(\vec{k}_{5}) | \Omega \rangle = {1 \over 2} \left( |\vec{k}_4, \vec{k}_5 \rin + |\vec{k}_4, \vec{k}_5 \rout \right) \\
&\langle \Omega | \spatinfann(\vec{k}_{1})\spatinfann(\vec{k}_{2}) = {1 \over 2} \left( \lout \vec{k}_1, \vec{k}_2|  +  \lin \vec{k}_1, \vec{k}_2 | \right).
\end{split}
\ee
Using CPT invariance (see the previous equation \eqref{cpt}), 
\be
	\begin{split}
		\Big[{M_5^{+}}\Big]_{\text{connected}}= \Big[
		\frac{1}{4}\Re({_\text{out}\langle k_1, k_2,k_3 | k_4,k_5 \rin})+\frac{1}{4} \Re({_\text{in}\langle k_1, k_2 |b_{\vec{k}_{3}}| k_4,k_5 \rin})\Big]_{\text{connected}},
		\\
		\Big[{M_5^{-}}\Big]_{\text{connected}}=\Big[\frac{1}{4}\Re({_\text{out}\langle k_1, k_2 |k_3, k_4,k_5 \rin})+\frac{1}{4} \Re({_\text{in}\langle k_1, k_2 |b^\dagger_{\vec{k}_{3}}| k_4,k_5 \rin})\Big]_{\text{connected}}.
	\end{split}
\ee
The first term in each line above is a $S$ matrix element. The second term can be written as a product of the $S$ matrix and its conjugate by inserting a complete set of \outip states. 
\be
	\begin{split}
		{}_{\text{in}}\braket{k_1,k_2|b_{\vec{k}_{3}}|k_4,k_5}_{\text{in}}\,&=\,\sum_{X} {}_{\text{in}}\braket{k_1,k_2|X}_{\text{out}}{}_{\text{out}}\braket{X|b_{\vec{k}_{3}}|k_4,k_5}_{\text{in}}.
	\end{split}
\ee
But the action of the \outip operators on \outip states is simple
\be
\lout X | b_{\vec{k}_3} = \lout \vec{k}_3, X |.
\ee
On the right, we have the state where we have added a single particle with momentum $\vec{k}_3$ to the excitations that are already present. 
This leads to
\be
\begin{split}
&\Big[{M_5^{+}}\Big]_{\text{connected}} = \frac{1}{4} \Re\Big(S_{{\vec{k}_1 \vec{k}_2} \vec{k}_3 \leftarrow \vec{k}_4, \vec{k}_5} + \sum_{X}(S_{{X} \leftarrow \vec{k}_1,\vec{k}_2  })^*S_{\vec{k}_3,X \leftarrow {\vec{k}_4,\vec{k}_5} }\Big) \\
 &\Big[{M_5^{-}}\Big]_{\text{connected}} = \frac{1}{4} \Re\Big(S_{\vec{k}_1 \vec{k}_2 \leftarrow  \vec{k}_3,\vec{k}_4, \vec{k}_5}+\sum_{X}(S_{{\vec{k}_3 ,X} \leftarrow \vec{k}_1,\vec{k}_2   })^*S_{{X}\leftarrow\vec{k}_4,\vec{k}_5   }\Big).
\end{split}
\ee
Although we have assumed that the $\vec{k}_i$ are distinct we still need to account for the trivial term in the S matrix. In terms of the T matrix the answer can be written as
\be
\label{vacuumfive}
\begin{split}
&\Big[{M_5^{+}}\Big]_{\text{connected}} = -\frac{1}{2} \Im\Big(T_{{\vec{k}_1 \vec{k}_2} \vec{k}_3 \leftarrow \vec{k}_4, \vec{k}_5} \Big)  + {1 \over 4} \Re \Big(\sum_{X}(T_{{X} \leftarrow \vec{k}_1,\vec{k}_2  })^*T_{\vec{k}_3,X \leftarrow {\vec{k}_4,\vec{k}_5} }\Big) \\
 &\Big[{M_5^{-}}\Big]_{\text{connected}} = -\frac{1}{2} \Im\Big(T_{\vec{k}_1 \vec{k}_2 \leftarrow  \vec{k}_3,\vec{k}_4, \vec{k}_5}\Big) + {1 \over 4} \Re \Big(\sum_{X}(T_{{\vec{k}_3 ,X} \leftarrow \vec{k}_1,\vec{k}_2   })^*T_{{X}\leftarrow\vec{k}_4,\vec{k}_5   }\Big).
\end{split}
\ee
Note the additional factor of 2 in the linear term.

As an example, consider the five-point correlator in a theory with the interaction term $\frac{\lambda}{3!}\phi^3+\frac{g}{4!}\phi^4$ to $\mathcal{O}(\lambda g)$. At this order, $X$ can range only over single-particle states. But the quadratic term cannot contribute because the three-point on-shell amplitude vanishes.   The linear terms in $T$, can in principle have contribution from all the $10$ channels. But the internal propagator cannot be put on shell consistent with energy-momentum conservation. For this reason, this term does not contribute either.  Therefore the $5$-point correlator vanishes in this theory to ${\cal O}(\lambda g)$. This can be verified by directly working out the on-shell part of the Wightman function and checking that different terms cancel precisely among each other.

To obtain a nonzero answer consider a different theory with the interaction term $\frac{\tilde{\lambda}}{2!}\phi^2 \chi+\frac{\tilde{g}}{3!}\phi^3\chi$, where $\chi$ is a scalar field of mass  $M > 2 m$. The external insertions are still taken to be the $\phi$ fields.  In this case, the internal propagator {\em can} go on shell. A perturbative calculation yields
\be \label{5 pnt vacuum}
	\begin{split}
		\langle \Omega | \momopspat(k_1) &\momopspat(k_2) \momopspat(k_3)\momopspat(k_4) \momopspat(k_5) | \Omega \rangle_{\text{connected}}
		\\& ={1 \over 4}\tilde{\lambda}\tilde{g}(2 \pi)^{4}\delta^4\Big(\sum_{i=1}^{5}k_i\Big)\Big(\prod_{i}(2 \pi)\delta(k_i^2+m^2)\Big)\theta(k_1^0)\theta(k_2^0)\theta(-k_4^0)\theta(-k_5^0)
		\\&
		\Big[\theta(k_3^0)\Big(-\D_{k_4k_5}-\D_{k_2k_3}-\D_{k_1k_3}\Big)
		+\theta(-k_3^0)\Big(-\D_{k_3k_5}-\D_{k_3k_4}-\D_{k_1k_2}\Big)\Big].
	\end{split}
\ee
Here $\D_{k_ak_b} \equiv 2 \pi \delta ((k_a+ k_b)^2+M^2))$.   Several other potential contributions cancel due to kinematic constraints as noted above. This matches precisely with the expression \eqref{vacuumfive}.
\subsection{$6$-point vacuum correlator }
Using precisely the same method as above, the $6$-point vacuum correlator can straightforwardly be written as a polynomial of the S matrix and its conjugate. We do not present the answer explicitly. 

Instead we present the perturbative answer for the  $6$-point correlator in a theory with the interaction $\frac{g}{4!}\phi^4$ to ${\cal O}(g^2)$. This correlator receives contributions from exchange diagrams. There are a total of $10$ different ``channels'' which can contribute to this correlator that are classified by which external legs connect to the two internal vertices. In addition, the internal vertices can run over the different folds of the contour. The final answer is remarkably simple and is given by
\be \label{6 pnt vacuum}
	\begin{split}
	&\langle \Omega | \momopspat(k_1) \momopspat(k_2) \momopspat(k_3)\momopspat(k_4) \momopspat(k_5)\momopspat(k_6)  | \Omega \rangle_{\text{connected}}
	\\& =\frac{g^2}{4}(2 \pi)^{4}\delta^4\Big(\sum_{i}^{6}k_i\Big)\Big(\prod_{i} (2 \pi) \delta(k_i^2+m^2)\Big)\theta(k_1^0)\theta(k_2^0)\theta(-k_5^0)\theta(-k_6^0)\theta(k_3^0)\theta(-k_4^0)
	\\& \quad \quad \quad \quad \quad \quad
	\Big(\D_{k_1k_2k_4}-\D_{k_1k_3k_5}-\D_{k_1k_3k_6}-\D_{k_1k_4k_5}-\D_{k_1k_4k_6}\Big),
\end{split}
\ee
 where, $\D_{k_ak_bk_c} \equiv 2 \pi \delta ((k_a+ k_b+k_c)^2+m^2))$. 
\subsection{$6$-point non-vacuum correlator}
As a final example, we study the amplitude with two insertions  each at $i^+, i^-,\spatinf$ in the same theory that was used in the previous subsection.
This amplitude again receives contributions from exchange diagrams. The final answer for the connected part is
\be \label{6 pnt nonvacuum}
	\begin{split}
		&\lout \vec{q}_1 ,\vec{q}_2 | \momopspat(k_1) \momopspat(k_2)   | \vec{p}_1,\vec{p}_2 \rin{}_{\text{,connected}}
		\\& =g^2(2 \pi)^{4}\delta^4\Big(\sum_{i=1}^{2}q_i+\sum_{i=1}^{2}k_i+\sum_{i=1}^{2}p_i\Big)\prod_{i=1}^{2}(2 \pi)\delta(k_i^2+m^2)
		\\&
		\Big[-{1 \over 2}\theta(k_1^0)\theta(k_2^0)\Big(T_{q_1q_2k_1}+T_{q_1q_2k_2}+T_{q_1q_2p_1}+T_{q_1 q_2 p_2}+T_{q_1k_1 k_2}+T_{q_1k_1p_1} + T_{q_1k_1p_2}
		\\& \qquad \qquad \qquad \qquad \qquad \qquad \qquad
+T_{q_1k_2p_2}+T_{q_1 k_2p_1}+T_{q_1p_1p_2}
		\Big)
		\\&
	-{1 \over 2}\theta(-k_1^0)\theta(-k_2^0)\Big(T_{q_1q_2k_1}+T_{q_1q_2k_2}+T_{q_1q_2p_1}+T_{q_1q_2p_2}+T_{q_1k_1k_2}+T_{q_1k_1p_1}+T_{q_1k_1p_2}
		\\& \qquad \qquad \quad  \qquad \qquad \quad   \qquad \quad  
		+T_{q_1k_2p_2}+T_{q_1k_2p_1}+T_{q_1p_1p_2}
		\Big)
		\\&
		-{1 \over 4}\theta(-k_1^0)\theta(k_2^0)\D_{q_1q_2k_1}-{1 \over 4}\theta(k_1^0)\theta(-k_2^0)\Big(\D_{q_1k_1p_1}+\D_{q_1k_1p_2}+\D_{q_1k_2p_1} + \D_{q_1k_2p_2}\Big)\Big].
\end{split}
\ee
Here, the four-momenta $p_1, p_2$ that appears on the right hand side are related to the three momenta that appear in the \inim state through $p_i = (-\omega_{\vec{p}_i}, -\vec{p}_i)$ corresponding to the standard sign-reversal that is carried out for incoming momenta. Also, $T_{k_lk_mk_n} = T(k_l+k_m+k_n) $ is the tree-level Feynman propagator and $\D_{k_lk_mk_n}$ is the same as above.

\section{Conclusion}
For the convenience of the reader, we start this section by summarizing all our main results.  Some parts of this paper are somewhat technical, and so the first subsection below repeats many of the important equations. We then discuss some possible future directions.

\subsection{Summary \label{secsummary}}

The dS$_3$ slice at infinite proper distance, $\rho$,  from an arbitrarily chosen origin is denoted by $\spatinf$. The authors of \cite{Laddha:2022nmj} proposed that a  bulk massive field $\phi(\rho, X)$ could be extrapolated to $\spatinf$ by stripping off an exponential tail. 
\be
\label{summarylimit}
\opspat(\smone)  = \lim_{\rho \rightarrow \infty}  \sqrt{\frac{2}{\pi}}\rho\,\sqrt{m \rho}  \, {e^{m \rho}} \int \phi(\rho, X) \smone(X) \dds. \qquad \text{(free theory)}
\ee
However, \cite{Laddha:2022nmj} showed that the field must be smeared with a function, $\smone(X)$ on $\spatinf$ with specific properties to obtain a boundary operator with finite fluctuations.

In this paper, we found a simple reformulation of this restriction. When the smearing function is decomposed in spherical harmonics,
\be
g_{\ell}(\tau) = \int \smone(\tau, \Omega) Y_{\ell}(\Omega) d^2 \Omega,
\ee
each $\smone_{\ell}(\tau)$ must be analytic in the range $\Im(\tau) \in (-{\pi \over 2}, {\pi \over 2})$ for $\opspat(\smone)$ to have finite fluctuations.

Given an on-shell four-momentum $k = (\omega_{k}, \vec{k})$, it is convenient to define the following transforms of a smearing function,
\be
\widetilde{\smone}^{\pm}(\vec{k}) \equiv \lim_{\rho \rightarrow \infty} D(\rho) \int e^{\pm i \rho k \cdot X} \smone(X) \dds = -{4 \pi \over m} \sum_{\ell}   \left( \smone_{\ell}(\tau_0 \mp i {\pi \over 2}) + (-1)^{\ell} \smone_{\ell}(-\tau_0 \mp i {\pi \over 2}) \right) Y^*_{\ell}(\hat{k}).
\ee
We showed that the two-point function in the free theory is compactly represented by
\be
\langle \Omega|  \opspat(\smone) \opspat(\smtwo) |\Omega \rangle = \int \widetilde{\smone}^{+}(\vec{k}) \widetilde{\smtwo}^{-}(\vec{k}) {d^3 \vec{k} \over (2 \pi)^3 2 \omega_k}. \qquad \text{(free~theory)}
\ee

In the free theory, we explicitly found an analogue of the HKLL-smearing function that represents bulk operators as smeared boundary operators. 
\be
\phi(t, \vec{x}) = {m \over \pi} \int d \tau' \cosh^2 \tau' d^2 \Omega' \opspat(\tau', \Omega') \sum_{\ell} k_{\ell}(m |\vec{x}| \cosh \tau') e^{m \sinh \tau' t} Y_{\ell}(\Omega') Y_{\ell}^*(\hat{x}).
\ee
The results above complete the description of free fields at $\spatinf$. All nonzero correlators factorize into two-point functions that we have analyzed comprehensively. 

We then turned to the behaviour of interacting fields at $\spatinf$. We showed that a naive use of equation  \eqref{summarylimit} does not yield sensible results. This is because, in the interacting theory, pushing the field to spatial infinity does not automatically put the field ``on shell''. The cure is to extract the on-shell part of the operator before extrapolating it.

Given the full Heisenberg operator in Fourier space $\phi(k)$, we extract the on-shell part of the operator through
\be
\spatinfann(\vec{k}) = \lim_{\epsilon \rightarrow 0} \int_{\omega_{k}-\epsilon}^{\omega_{k}+\epsilon} \phi(k) {(2 |k^0|) d k^0 \over 2 \pi}; \qquad \spatinfcre(-\vec{k}) = \lim_{\epsilon \rightarrow 0} \int_{-\omega_{k}-\epsilon}^{-\omega_{k}+\epsilon} \phi(k) {(2 |k^0|) d k^0 \over 2 \pi}.
\ee
These on-shell parts can be reassembled into a field,
\be
\phi_{\spatinf}(x)= \int {d^3 \vec{k} \over (2 \pi)^3 2 \omega_{k}} \left( \spatinfann(\vec{k}) e^{i k \cdot x} + \spatinfcre(\vec{k}) e^{-i k \cdot x} \right).
\ee

In the free field theory, $\phi_{\spatinf}(x)$ coincides with the field itself. In the interacting theory, it is that part of the full Heisenberg operator which continues to be concentrated on the mass shell. The generalization of \eqref{summarylimit} to the interacting theory is simply
\be
\opspat(\smone)  = \lim_{\rho \rightarrow \infty}  \sqrt{\frac{2}{\pi}}\rho\,\sqrt{m \rho}  \, {e^{m \rho}} \int \phi_{\spatinf}(\rho, X) \smone(X) \dds, 
\ee
which reduces to \eqref{summarylimit} in the free limit.

The boundary operators become
\be
\opspat(\smone) =  \int {d^3 \vec{k} \over (2 \pi)^3 2 \omega_{k}} \left(\spatinfann(\vec{k}) \widetilde{\smone}^{+}(\vec{k})  + \spatinfcre(\vec{k}) \widetilde{\smone}^{-}(\vec{k}) \right).
\ee
This formula can be used to translate between correlators of $\spatinfop(\vec{k})$ and boundary correlators for any smearing function.
 
We showed that the $\spatinfop({\vec{k}})$ operators have a simple relation with the conventional \inim operators $a_{\vec{k}}, a^{\dagger}_{\vec{k}}$ that are defined at $i^{-}$ and the \outip operators $b_{\vec{k}}, b^{\dagger}_{\vec{k}}$ that are defined at $i^{+}$. 
\be
\label{phipmabsummary}
\spatinfann({\vec{k}}) = {1 \over 2}(a_{\vec{k}} + b_{\vec{k}}); \qquad \spatinfcre(\vec{k}) = {1 \over 2} (a_{\vec{k}}^{\dagger} + b_{\vec{k}}^{\dagger}).
\ee
The $a$ and $b$ operators do not have simple commutation relations because $b_{\vec{k}} = S^{\dagger} a_{\vec{k}} S$ where $S$ is the S-matrix. 

An equivalent way of defining boundary correlators is as follows. Given the full Wightman function between arbitrary external states, we extract the part that is on shell.
\be
\langle \Psi_1 | \phi(k_1) \ldots \phi(k_n) | \Psi_2 \rangle = W^{\Psi_1, \Psi_2}_{\spatinf}(k_1,\ldots,k_n) + \ldots,
\ee
where 
\be
W^{\Psi_1, \Psi_2}_{\spatinf}(k_1,\ldots,k_n) = G^{\Psi_1, \Psi_2}(k_1,\ldots,k_n) \prod_{i=1}^{n} (2 \pi) \delta(k_i^2 + m^2),
\ee
 is the term in the full Wightman correlator that contains a mass-shell delta function for {\em each} external momentum and the $\ldots$ are the remaining part of the Wightman correlator.  The LSZ formula tells us that the S-matrix is given by computing the bulk time-ordered Green's function and focusing on the residue of the pole on mass-shell for each external momentum. The formula above provides a natural generalization: the correlator relevant for spatial infinity is given by computing the bulk Wightman function and focusing on the part that contains a mass-shell delta function for each external momentum.

The Fourier transform of this function is the correlator of the on-shell operators defined above.
\be
W^{\Psi_1, \Psi_2}_{\spatinf}(x_1,\ldots,x_n) =  \langle \Psi_1 | \phi_{\spatinf}(x_1),\ldots,\phi_{\spatinf}(x_n) | \Psi_2 \rangle.
\ee

In section \ref{secsample}, we provide several examples of perturbative correlators at $\spatinf$. We also describe correlators with insertions at $i^{-}, i^{+}$ and $\spatinf$. All these correlators can be computed in two ways. The first is to simply compute the interacting Wightman function and extract the part that is proportional to delta functions in the external momenta. Perturbation theory for Wightman functions is straightforward but tedious, as reviewed in Appendix \ref{reviewfeynman}. Alternately, we can use the relation \eqref{phipmabsummary} to convert the correlator of on-shell operators to the S-matrix and its powers and then use standard time-ordered perturbation theory to compute the S matrix. 

\subsection{Discussion}
In this paper, we have described a novel limit for interacting massive fields at spatial infinity. The motivation for this study comes from the principle of holography of information, which states that bulk observables in a gravitational theory should permit another description at the boundary of a Cauchy slice. However, in this paper, our analysis was restricted to nongravitational quantum field theories in four-dimensional Minkowski space. Therefore, our results constitute steps in a larger program, and do not have any immediate implications for flat-space holography.

It is clearly of interest to generalize these techniques to {\em asymptotically} flat space. This would put us in a position to ask questions about theories with dynamical gravity. It is also of interest to understand whether massless fields can be described from spatial infinity instead of null infinity, which would be necessary if we seek a unified description of gravitational theories at spatial infinity.

It is conventional to describe boundary observables in flat space on $i^{\pm}$ and ${\cal I}^{\pm}$.  However, viewed individually, these algebras are all free: the S-matrix is nontrivial because it involves cross correlators between operators defined on the past boundaries with those defined on future boundaries. In contrast, correlators at $\spatinf$ capture interactions on a single boundary. This is reminiscent of AdS, where the boundary operator algebra retains information about interactions.

In the spirit of the AdS conformal bootstrap  \cite{Poland:2018epd}, one could ask whether it is possible to place general constraints on correlators at $\spatinf$ and then use them to derive consequences about the bulk theory. It is important to note that our extrapolation procedure breaks conformal invariance. So, while flat-space amplitudes display remarkable structure \cite{Prabhu:2023oxv}, AdS techniques must be suitably adapted before they can be applied to flat space.

We would like to understand the relationship of our work to celestial holography \cite{Fan:2020xjj,Albayrak:2020saa,Law:2020xcf,Raclariu:2021zjz,McLoughlin:2022ljp,Pasterski:2021raf,Donnay:2022aba,Donnay:2021wrk}.  How are celestial amplitudes related to correlators on $\spatinf$?  Although the literature commonly refers to the ``celestial sphere'', celestial amplitudes are computed as integral transformations of scattering amplitudes, which are defined on the entire conformal boundary. Therefore, it is not clear to us where the celestial sphere lives. Perhaps work on the holography of information could help to address this question.

Recently, \cite{Jain:2023fxc} showed that the S matrix can be derived by differentiating by taking functional derivatives of the path integral performed with past and future boundary conditions.  A natural question is whether correlators at $\spatinf$ can be understood similarly. 

\paragraph{\bf Acknowledgments} We are grateful to Tuneer Chakraborty, Chandramouli Chowdhury, Abhijit Gadde,  Diksha Jain, Alok Laddha,  R. Loganayagam,  Shiraz Minwalla, Siddharth Prabhu, Omkar Shetye and Pushkal Shrivastava  for helpful discussions and to Sumer Pingle Raju for typographical assistance.  S.R. was partially supported by a Swarnajayanti fellowship, DST/SJF/PSA-02/2016-17, of the Department of Science and Technology while this work was in progress.  Preliminary results from this work were presented at TIFR (Mumbai), University of Amsterdam, Stanford University,  Caltech, Kavli Institute for Theoretical Physics (KITP)  and at the Indian Strings Meeting 2023 at IIT Bombay. We are grateful for the discussions at these talks and the hospitality provided by these institutions. Through KITP, this research was supported in part by grant NSF PHY-2309135.   Research at ICTS-TIFR is supported by the Department of Atomic Energy, Government of India, under Project Identification Nos. RTI4001.

\appendix

\section*{Appendix}

\section{Perturbation theory for Wightman functions}\label{reviewfeynman}

In this section we rederive Feynman rules of the perturbation theory for Wightman function. These were derived in \cite{Ostendorf:1982uu}, \cite{steinmann1993perturbation} . For a simpler derivation in context of thermal correlation function see \cite{Banerjee:2019kjh} which we elaborate upon here. We also refer the reader to \cite{Chaudhuri:2018ihk,Haehl:2017eob,Chaudhuri:2018ymp} for a discussion of several aspects of out-of-time-ordered correlation functions.

\begin{figure}[H]
	\centering
	\hspace{1.7cm}
	\includegraphics[scale=0.3]{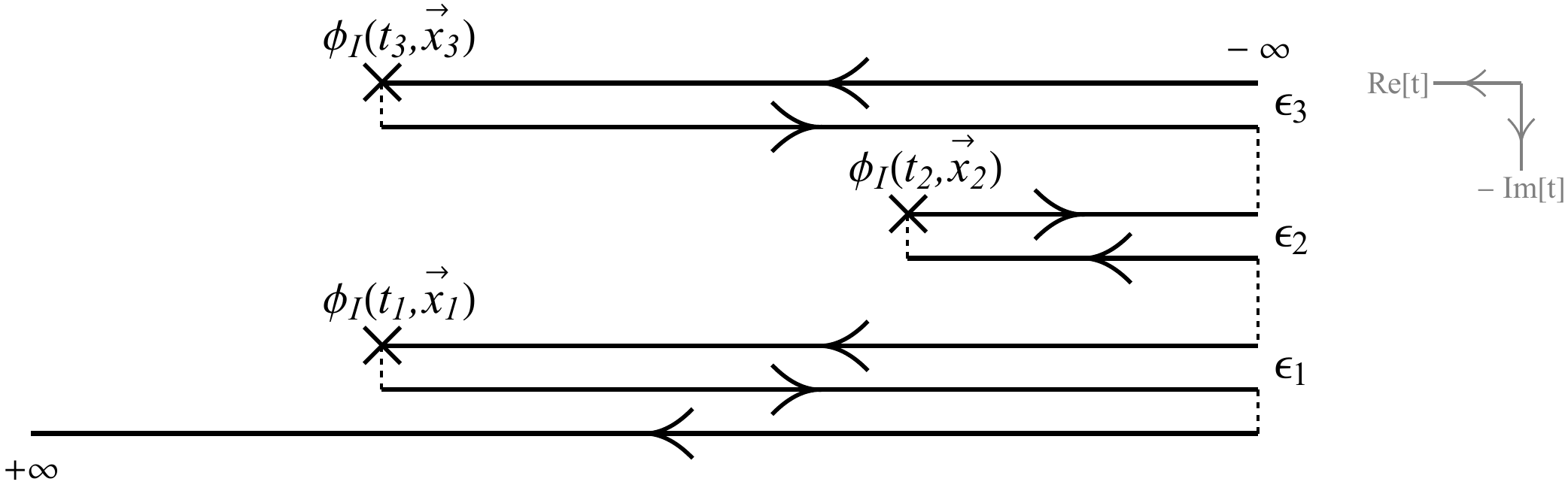}
	\caption{\em Contour for a 3 point function \label{figinit3pt}}
\end{figure}

We want to calculate an n-point Wightman correlator ${W}(x_1,x_2,\ldots x_n)$ in the interacting vacuum $\ket{\Omega}$.
\be
	\begin{split}
	{W}(x_1,x_2,\ldots x_n)\equiv&\mel{\Omega}{\phi(x_1)\phi(x_2)\ldots \phi(x_n)}{\Omega}.
\end{split}
\ee
The operators are {\em not} time ordered; the ordering is picked by hand, as displayed above.  It is convenient to extend the time variables of all the insertions in the complex plane by adding small imaginary parts
\be
t_i \rightarrow t_i - i \epsilon_i,
\ee
where we maintain the ordering $\epsilon_1 > \epsilon_2> \ldots> \epsilon_n~$.
The Wightman correlator is analytic under this extension, provided the imaginary parts are in the same order as the insertions. To obtain a perturbative expansion we convert the Heisenberg field $\phi(t_i, \vec{x}_i)$ to the interaction-picture field $\phi_I^{(i)} \equiv \phi_I(t_i,\vec{x}_i)$.
\be
		\phi(t_i, \vec{x}_i) = U(t_0, t_i)\phi_I^{(i)}U(t_i,t_0),
\ee
where $U$ is the interaction picture time evolution operator  \cite{peskin1995iqf} and $t_0$ is a reference time that is used to define the interaction picture. From a standard analysis we get the following relation between the interacting vacuum and the free vacuum
\be
	\ket{\Omega} = \lim_{T \rightarrow \infty(1 - i \epsilon)} \frac{1}{\braket{\Omega|0}}U(t_0,-T)\ket{0}  \quad; \quad     \bra{\Omega} = \lim_{T \rightarrow \infty (1 - i \epsilon)} \frac{1}{\braket{0|\Omega}} \bra{0}U(T,t_0)~,
\ee
where we have set the ground state energy to 0.  With this we can write 
\be
	\begin{split}
		&{W}(x_1,x_2,\ldots, x_n)\\
		=& \lim_{T \rightarrow\infty(1-i\epsilon)} \frac{1}{\mel{0}{U(T,-T)}{0}}\\
		\times & \mel{0}{U(T,t_0)U (t_0,t_1)\phi_{I }^{(1)} U(t_1,t_0)  \ldots U (t_0,t_n)\phi_{I }^{(n)} U(t_n,t_0) U(t_0,-T)}{0}.
\end{split}
\ee
The sequence of time evolutions can be thought of as going back and forth on a complexified contour as depicted in Figure \ref{figinit3pt}

For further simplification we follow the steps below. 
\begin{enumerate}
	\item We extend all the contour lines from $t= - \infty$ to $t= \infty$ by inserting the identity operator $1 =  U(t_i-\epsilon_{i},T-\epsilon_{i})U(T-\epsilon_{i},t_i-\epsilon_{i})$. This is shown in Figure \ref{figextension}
		\begin{figure}[H]
		\centering
		\hspace{1.7cm}
		\includegraphics[scale=0.3]{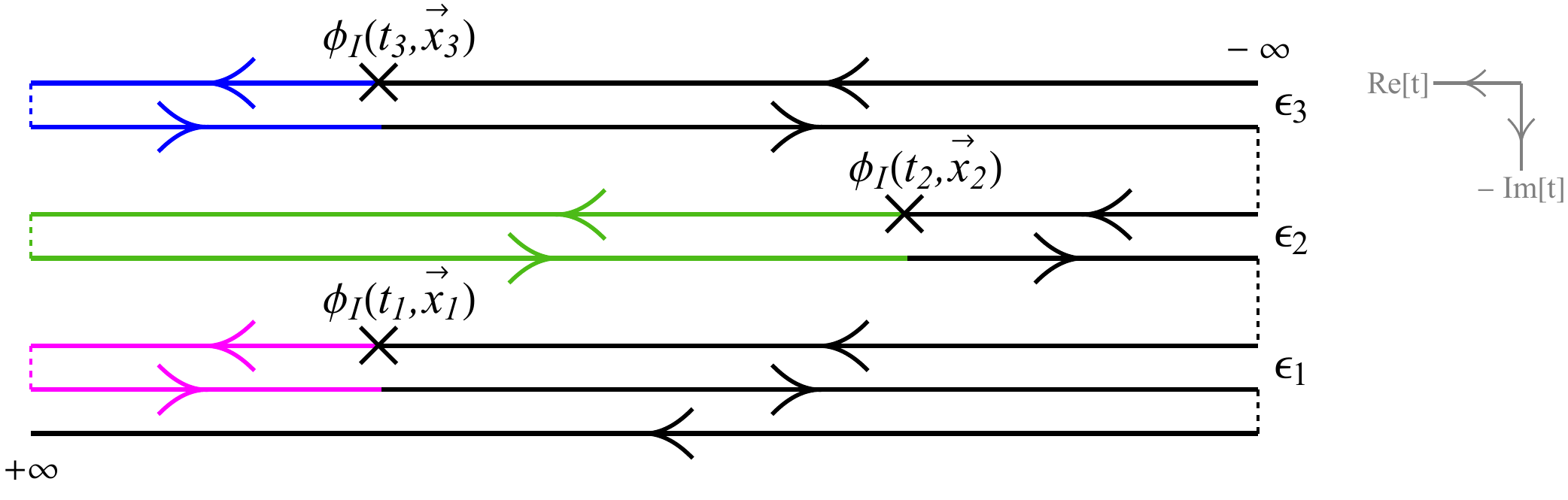}
		\caption{\em We extend contour lines from $- \infty$ to $\infty$ for a 3 point function. The extended segments that have been added to the original contour are colored  \label{figextension}}
	\end{figure}
	\item 
          We then  collapse the contour by deleting segments that cancel between each other to get a simplified final contour. This is shown in Figure \ref{figcollapse}
		\begin{figure}[H]
		\centering
		\hspace{1.7cm}
		\includegraphics[scale=0.3]{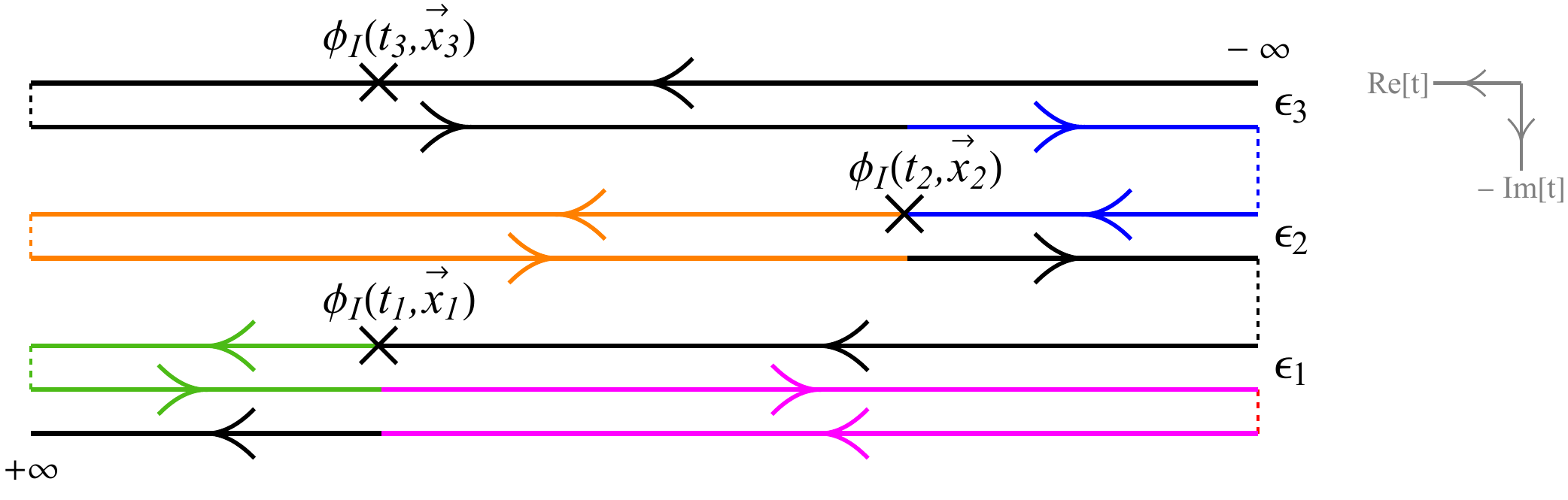}
		\caption{\em We collapse the contour lines by deleting segments that cancel mutually. The segments that will be deleted are colored. \label{figcollapse}.}
	\end{figure}
	For an odd number of external points, the collapsed contour represents the expression 
	\be
		\begin{split}
			&{W}(x_1,x_2,\ldots , x_n)\\
			=      & \lim_{T \rightarrow \infty(1- i \epsilon )}\frac{1}{{\mel{0}{U(T,-T)}{0}}}\\			
			\times & {\mel{0}{U(T,t_1)\phi_I^{(1)} U(t_1,-T)U(-T,t_2)\phi_I^{(2)}U(t_2,T) \ldots U(T,t_n)\phi_I^{(n)}U(t_n,-T)}{0}}.
\end{split}
	\ee
	\begin{figure}[H]
		\centering
		\hspace{1.7cm}
		\includegraphics[scale=0.3]{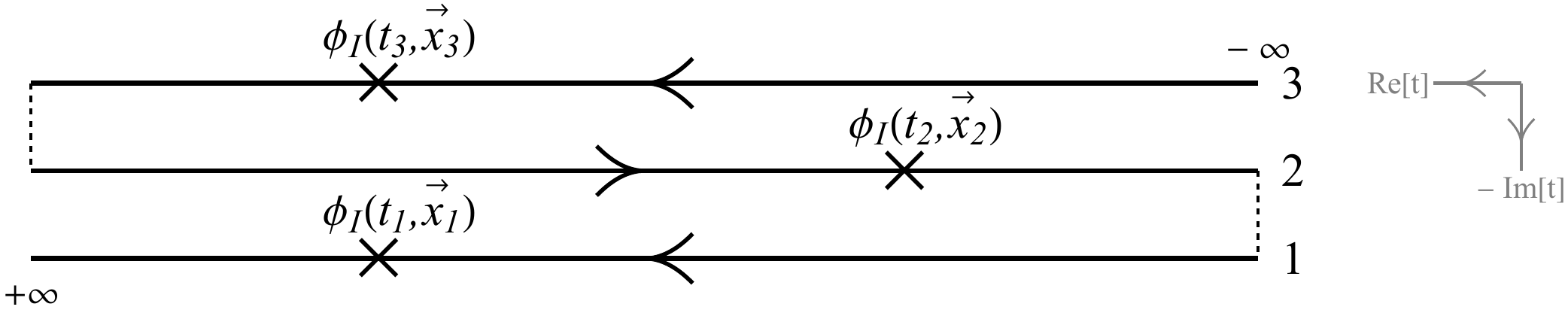}
		\caption{\em Collapsed contour for a 3 point function}
		\label{3 point}
	\end{figure}
	For an even number of external points, the collapsed contour represents the expression
	\be
		\begin{split}
			&{W}(x_1,x_2,\ldots , x_n)\\
			=      & \lim_{T \rightarrow \infty(1- i \epsilon )}\frac{1}{{\mel{0}{U(T,-T)}{0}}}\\			
			\times & {\mel{0}{U(T,-T)U(-T,t_1)\phi_I^{(1)} U(t_1,T)U(T,t_2)\phi_I^{(2)}U(t_2,-T) \ldots U(T,t_n)\phi_I^{(n)}U(t_n,-T)}{0}}.
\end{split}
	\ee
\begin{figure}[H]
	\centering
	\hspace{1.7cm}
	\includegraphics[scale=0.3]{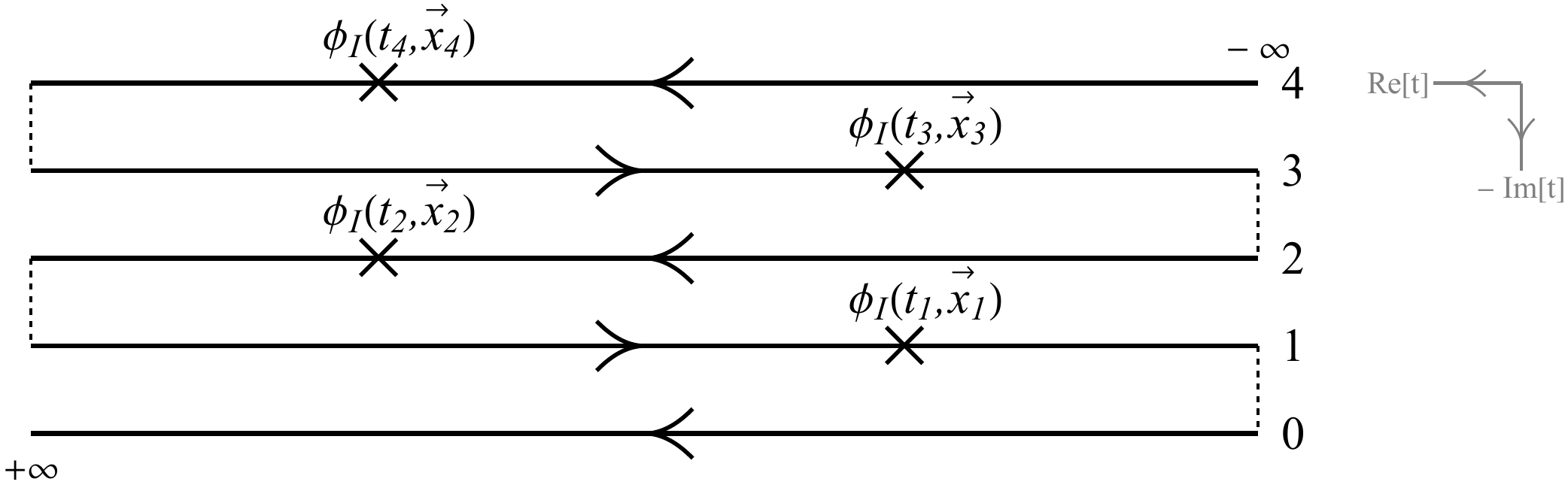}
	\caption{\em Collapsed contour lines for a 4 point function}
	\label{4 point}
\end{figure}
We find there is an extra contour line running from $t = - \infty$ to $t = \infty$ when the number of  external points is even. 
For both the cases the correlators can be written as contour ordered product. 
	\be
		\begin{split}
			{W}(x_1,x_2,\ldots x_n)  = & \lim_{T \rightarrow \infty(1- i \epsilon )} \frac{\mel{0}{ \mathcal{T}_{\mathcal{C}} \Big\{\phi_I(x_1)\phi_I(x_2)\ldots \phi_I(x_n) e^{- i \int_{-T}^{T} dt H_I (t)}\Big\}}{0}}{\mel{0}{\mathcal{T}_{\mathcal{C}}\Big\{e^{- i \int_{-T}^{T} dt H_I (t)}\Big\}}{0}}.
\end{split}
	\ee
In the above expression the symbol $\mathcal{T}_C$ denotes contour ordering on the final collapsed contour obtained using the manipulations above. We note that the time-ordering symbol for time-ordered correlators has been altered to contour ordered symbol for Wightman functions. 
\end{enumerate}
\subsection{Momentum space Feynman rules}
We can go to momentum space by taking the Fourier transform.
\be
	\begin{split}
	{W}(x_1,x_2,\ldots x_n)= \int \prod _{i = 1}^{n} \frac{d^{4}{k_i}}{(2 \pi )^4}  e^{ \sum_{i=1}^{n}{i k_i.x_i}}  \underbrace{{W}(k_1,k_2,\ldots,k_n)  }_{\substack{\text{{\small{Momentum space  }}}\\ \text{{\small{ Wightman function}}}}}.
\end{split}
\ee
After expanding the time-evolution operators perturbatively and applying Wick's theorem we arrive at the following contour prescription and Feynman rules. 

\paragraph{{\bf Contour  prescription. }}
The number of contour lines depends on whether the number of external points $n$ is odd or even. If $n$ is odd, the number of contour lines is $n$; whereas if $n$ is even the number of contour lines is $(n+1)$. We place $i$th external point at $i$th contour. For $n$ even, we draw an extra contour line (labelling it as the $0^{\text{th}}$ line) which runs from $- \infty$ to $\infty$.
\paragraph{\bf{Feynman rules.}}
\begin{enumerate}
\item The interaction term can be on any of the $n$ contours. If the interaction term is on the $i$th contour with $(n-i) $ even, then we add the interaction vertex  $-i H_{I}$ where $H_{I}$ is the interaction Hamiltonian.                     If $(n-i)$ is odd we add the interaction  vertex $i H_{I}$
		\item We can have a propagator joining a point on the i\textsuperscript{th} contour line to another point on the j\textsuperscript{th} contour line with the momentum $k$  flowing  from $j$ to $i$.
		\begin{figure}[H]
			\centering
			\hspace{1.7cm}
			\includegraphics[scale=0.15]{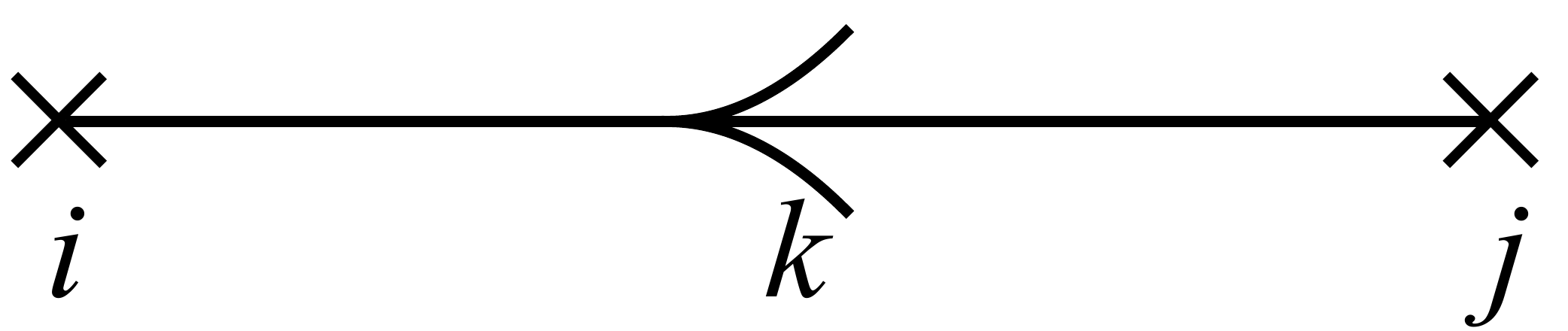}
			\caption{Convention for propagator $D_{ij}(k)$}
		\end{figure}
		
		 For different $i,j$ we have the following prescription. 
				\be\label{rules}
			D_{ij}(k)\,=\,
			\begin{cases} 
				T(k)=-\dfrac{i}{k^{2}\,+\,m^{2}\,-\,i \epsilon} & i\,=\,j \  \text{and} \  (n-i) \ \text{even}, \\
				\overline{T}(k)=\dfrac{i}{k^{2}\,+\,m^{2}\,+\,i \epsilon} & i\,=\,j \ \text{and} \ (n-i)\ \text{odd}, \\
			W(k)=2 \pi \theta(k^{0})\,\delta(k^{2}+m^{2}) & i\,<\,j, \\
			\overline{W}(k)=2 \pi \theta(-k^{0})\,\delta(k^{2}+m^{2}) & i\,>\,j.
			\end{cases}
\ee
\item
We impose energy-momentum conservation at each vertex as usual and integrate over undetermined momenta.
\end{enumerate}
Note that, unlike the usual time-ordered Feynman rules, here we find different kinds of propagators and the signs of the interaction Hamiltonian change according to where it is inserted.

\paragraph{\bf Spectral condition.}
Quantum field theories obey a spectral condition. In any energy-momentum eigenstate, the energy must be greater than the magnitude of the momentum $E > |\vec{P}|$. This leads to some simple spectral conditions on Wightman functions. Momentum space Wightman functions $W(k_1,k_2,\ldots,k_n)$ are only supported over the following domain.
\be
	K_m  =\sum_{i}^{m}k_i\in V^+ \quad\text{for all $m<n$ where}\quad  V^+ = \{ k ; k^0 >0 ; k^2<0\}.
\ee 
These spectral conditions are obeyed by Wightman functions computed according to the rules above \cite{steinmann1993perturbation} although this might be nontrivial to check.

\subsection{Renormalization}
In any interacting quantum field theory, we expect the propagators to get renormalized. We emphasize that the mass that has appeared everywhere in the paper is the {\em physical mass} of the field and not the bare mass that enters the Lagrangian. We recall that time-ordered Green's functions have poles at the physical mass and the LSZ formula relates the residue to the S matrix; Wightman functions have delta functions at the physical mass and our prescription relates this on-shell part to the correlator at $\spatinf$. 

In perturbation theory, we can resum corrections to the bare propagators displayed above to obtain renormalized propagators. 
\be
\label{resummedprops}
\begin{split}
&T(k)=\int e^{-i k \cdot (x - y)} \langle \Omega | {\cal T}\{\phi(x) \phi(y)\} | \Omega \rangle; \qquad \overline{T}(k)=\int e^{-i k \cdot (x - y)} \langle \Omega | {\overline{\cal T}}\{\phi(x) \phi(y)\} | \Omega \rangle;  \\
&W(k)=\int e^{-i k \cdot (x - y)} \langle \Omega | \phi(x) \phi(y) | \Omega \rangle; \qquad \overline{W}(k)=\int e^{-i k \cdot (x - y)} \langle \Omega | \phi(y) \phi(x) | \Omega \rangle.
\end{split}
\ee
Note that $W(k) = \theta(k^0) \rho(k^2)$ where $\rho(k^2)$ is the ``spectral density'' that appears in the Kallen-Lehmann spectral representation.

After renormalization, we expect 
\begin{enumerate}
\item
$T(k)$ has a pole at the physical mass $T(k) = {-i \over k^2 + m^2 - i \epsilon} + \ldots$
\item
$\overline{T}(k)$ has a pole at the physical mass ${i \over k^2 + m^2 + i \epsilon} + \ldots$.
\item
$W(k)$ has a delta function at the physical mass, $W(k) = \theta(k^0) \left( \delta(k^2 + m^2) + \ldots\right)$
\item
$\overline{W}(k)$ has a delta function at the physical mass, $\overline{W}(k) = \theta(-k^0) \left(\delta(k^2 + m^2) + \ldots \right)$
\end{enumerate}
where in each case the $\ldots$ denote functions of $k^2$ that do not have singularities at $k^2+m^2 = 0$. Note that, for simplicity, we focus on a renormalization scheme where there are no wave-function renormalization factors. Correspondingly, the field is normalized so that its matrix elements between one-particle states and the vacuum are
\be
\label{normappendix}
\langle k | \phi(x) |\Omega \rangle = e^{-i k \cdot x}; \qquad \langle \Omega | \phi(x) |k \rangle = e^{i k \cdot x}.
\ee

\begin{figure}[H]
    \centering
~     \begin{subfigure}[t]{0.48\textwidth}
        \centering
	\includegraphics[width=\textwidth]{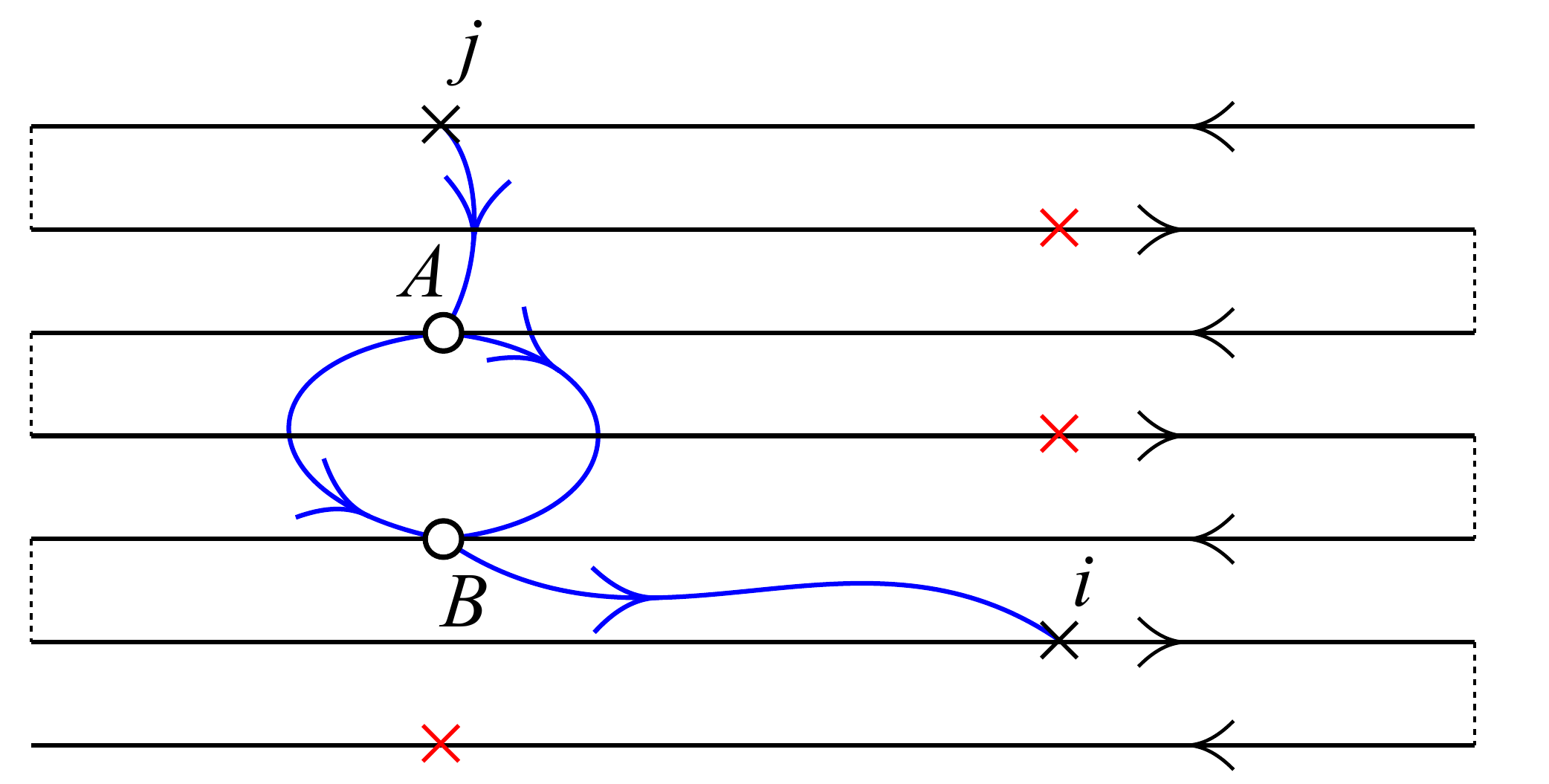}
\caption{}
	\label{blob}
    \end{subfigure}
~
        \begin{subfigure}[t]{0.48\textwidth}
        \centering
	\includegraphics[width=\textwidth]{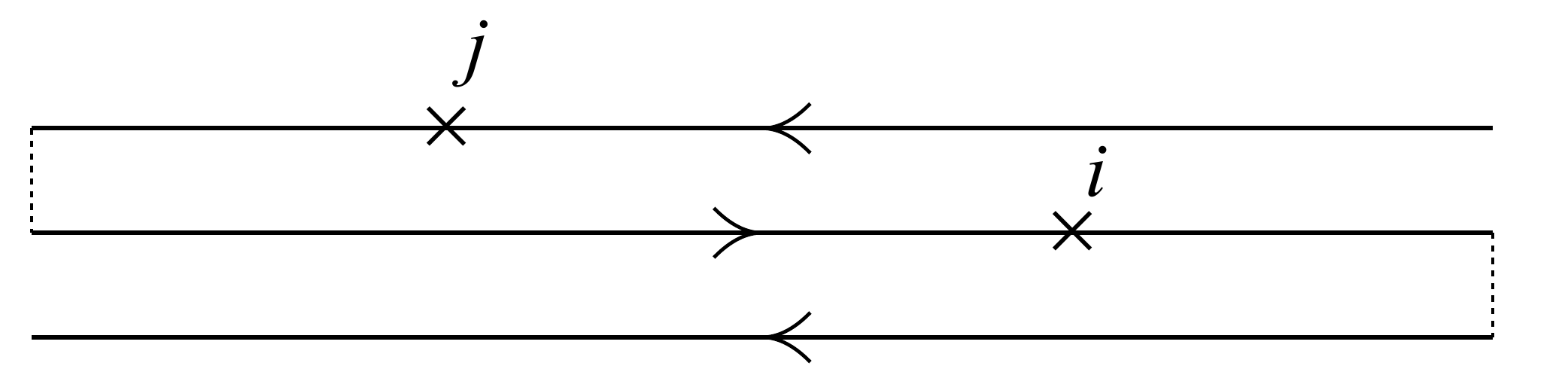}
	 \caption{}
\label{figwightrenorm}
\end{subfigure}
\caption{A ``self-energy'' type correction to a Wightman propagator might seem to depend on the number of folds in the contour and the presence of other vertices (red) as shown on the left. However, a contour collapsing argument shows that it is always sufficient to consider only the $3$ contour folds shown on the right.
}
\end{figure}

The resummed time-ordered and anti-time-ordered propagators can evidently be computed by summing all corrections to the two-point function on a single fold of the contour. In perturbation theory, the Wightman (or anti-Wightman) propagator might be used to connect two vertices on distant folds of a multi-fold contour.   One might worry that the corrections to the propagator depend on the number of folds in the contour. (See Figure \ref{blob}.) However, a contour collapsing argument shows that to study corrections to the propagator, one can ignore the rest of the diagram and simply focus on a three-fold contour displayed in Figure \ref{figwightrenorm}.

\subsection{Wightman function for non-vacuum states}
\begin{figure}[H]
 	\centering
 	\hspace{1.7cm}
 	\includegraphics[scale=0.3]{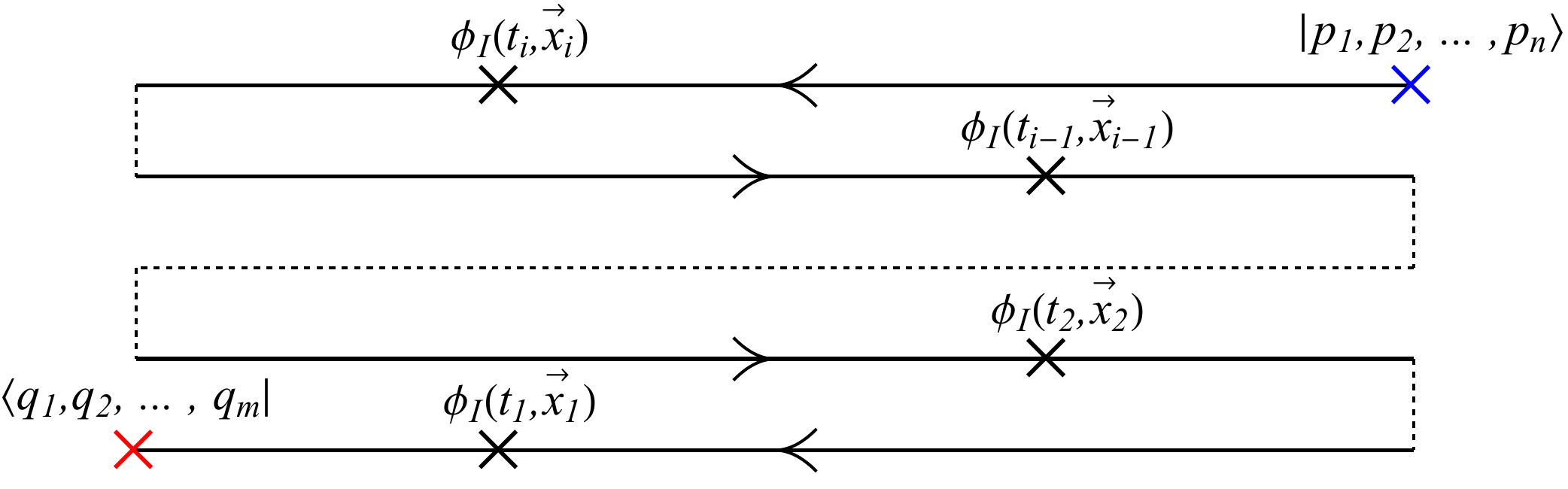}
 	\caption{\em Contour lines with external states placed on the ends of the contour.}
 	\label{6point_222revfeyn}
 \end{figure}

We also calculate
 \be
W^{\Psi_1, \Psi_2}(x_1,\ldots,x_n) = \langle \Psi_1 | \phi(x_1) \ldots \phi(x_n) | \Psi_2 \rangle,
\ee
when $\Psi_1,\Psi_2$ are non-vacuum states. By convention, we take the states on the right to be \inim states and the states on the left to be \outip states: they take the form ${}_{\text{out}}\bra{\vec{q}_1,\vec{q}_2,\ldots ,\vec{q}_{n}}$ and $\ket{\vec{p}_1,\vec{p}_2,\ldots ,\vec{p}_{m}}_{\text{in}}$  Since the \inim states are defined at $i^-$ we can think of them as prepared by insertions placed at the top of the contour on the extreme right; the \outip states  are prepared at $i^+$ and so we can think of them as prepared by insertions placed at the bottom of the contour on the extreme left. See Figure \ref{6point_222revfeyn} for an illustration.

Two points must be kept in mind
\begin{enumerate}
\item
It is convenient to adopt the standard convention that the four-momentum that enters the Feynman rules for \inim states is related to the physical momentum after a sign flip.
\be
p = (-\omega_{\vec{p}}, -\vec{p}).
\ee
\item
We must amputate the propagators that connect internal vertices to the external states. This can be thought of as the pair of Feynman rules
 \be\label{rules_non-vacuum}
\lout \wick{\c{\vec{q}_i} | \c{\phi} (k)} = (2 \pi)^4 \delta^4(q_i + k);
\quad \wick {\c{\phi}(k) \c{{|\vec{p}_i}} }  \rin = (2 \pi)^4 \delta^4(p_i + k).
\ee
Note that $p_i^0 < 0$ and $q_i^0 > 0$ with the convention above. 
\end{enumerate}

\subsection{Insertions at $\spatinf$}
Finally, we turn to the case where some of the external points are at $\hat{i}^{0}$. Most of the Feynman rules above are unchanged except that the propagators that connect internal vertices to external states must be truncated as follows. 
\be\label{modifiedrules}
	D^{\text{external}}_{ij}(k)\,=\,
	\begin{cases} 
		\TI(k)= \pi \delta(k^2+m^2)& i\,=\,j \  \text{and} \  (n-i) \ \text{even}, \\
	\ATI(k)= \pi \delta(k^2+m^2)& i\,=\,j \ \text{and} \ (n-i)\ \text{odd}, \\
		\w(k)=2 \pi \theta(k^{0})\,\delta(k^{2}+m^{2}) & i\,<\,j, \\
	\wb(k)=2 \pi \theta(-k^{0})\,\delta(k^{2}+m^{2}) & i\,>\,j.
	\end{cases}
\ee
This is because the prescription of section \ref{seconshellinteract} instructs us to focus on the piece of the Wightman function that has delta functions on mass shell for each external operator that will be extrapolated to $\spatinf$. These delta functions can come only from the external propagators. 

\paragraph{\bf External propagators are not renormalized.}
We remind the reader of an important point. In renormalized perturbation theory, it is convenient to resum the propagator and replace the internal propagators displayed in \eqref{rules} with the corrected propagators \eqref{resummedprops}. However, the external propagators are {\em not} renormalized. For external points at $i^+$ and $i^-$ we continue to use \eqref{rules_non-vacuum} and for points at $\spatinf$ we continue to use \eqref{modifiedrules}. 

To emphasize this, we use different symbols for the internal Wightman propagator $W(k)$ and the external Wightman propagator $\w(k)$ even though they coincide at tree level.

\section{Details of computations}
\label{details_of_comp}
In this section we will compute  Wightman correlators in massive scalar theories with the interaction Hamiltonian $ \frac{\lambda}{3!}\phi^{3}$ and $ \frac{g}{4!}\phi^{4}$ from Wightman perturbation theory. We will compute vacuum correlators as well as correlators with non-trivial \inim and \outip states. 

We remind the reader of a useful identity
\be
	T(k)+\overline{{T}}(k)\,=\,W(k)+\overline{W}(k).
\ee
Note that the above identity holds true when $T$ is replaced with $\TI$ and $W$ is replaced with $\w$.

A Mathematica notebook with details of these computations is available upon request.
\subsection{Vacuum Wightman correlators}
\subsubsection{3-point correlator \label{appsubsecthreept}}
In this subsection we will compute the $3-$point correlator $W_{\spatinf}(k_1, k_2, k_3)$ with the interaction Hamiltonian $\frac{\lambda^3}{3!}\phi^{3}$. We are interested in the regime $k_1^0>0, k_3^0<0$ since the correlator vanishes otherwise by the spectrum condition. 
By the argument in section \ref{subsecvanish}, we expect this correlator to vanish at $\Or[\lambda]$. The point of the computation below is to show that how this cancellation occurs in Wightman perturbation theory. 
\begin{figure}[h]
	\centering
	\begin{subfigure}[b]{0.6\textwidth}
		\includegraphics[width=\textwidth]{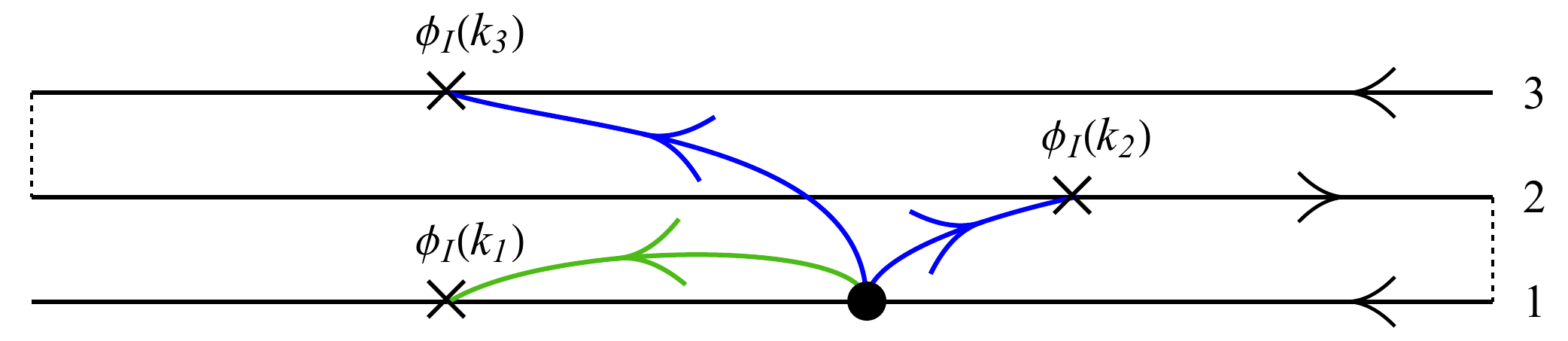}
		\caption{}
	\end{subfigure}
	
	\begin{subfigure}[b]{0.6\textwidth}
		\includegraphics[width=\textwidth]{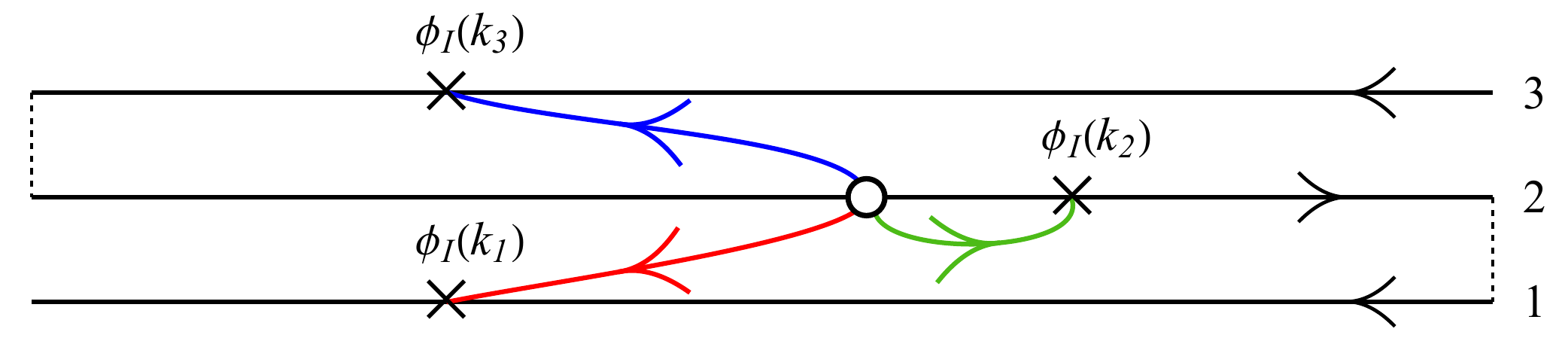}
		\caption{}
	\end{subfigure}
	
	\begin{subfigure}[b]{0.6\textwidth}
		\includegraphics[width=\textwidth]{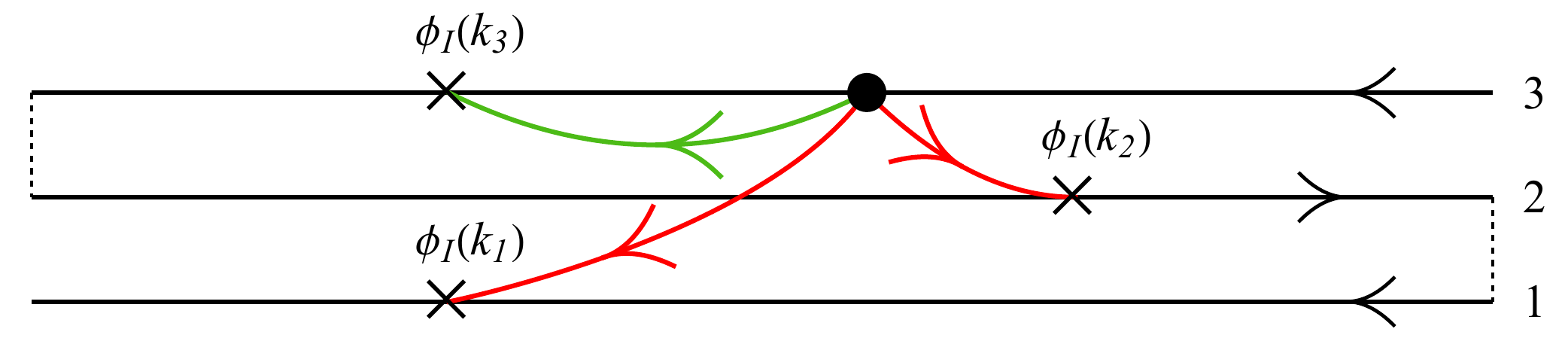}
		\caption{}
	\end{subfigure}
	
	\caption{We illustrate contour diagrams for 3 point contact interaction. There are total 3 diagrams. Diagrams (a),(b),(c) correspond to configurations where interaction point is at contour $1,2,3$ respectively. Red (blue) arrows correspond to (anti-)Wightman propagator, whereas green arrows correspond to modified time-ordered \& anti-time-ordered propagators given by \eqref{modifiedrules}. The filled circle corresponds to the vertex factor $-i \lambda$ and the unfilled circle to the vertex factor $i \lambda$.}
	\label{3_contour}
\end{figure}

To compute this 3-point correlator we need to consider the contour given in figure \ref{3_contour}. The internal point will run over all the contour lines. If the two points connecting the propagators are on the same contour line, we will get a time-ordered or an anti-time ordered propagators  and when these points are on different contour lines, we will get Wightman or anti- Wightman propagators. Now, by applying the rules \eqref{rules}, and \eqref{modifiedrules}, we find
\be
	\begin{split}
&{{W}_{\spatinf \text{contact}}(k_1,k_2,k_3) }\,
\\&=\,i \lambda\,(2 \pi)^4\delta^4\Big(\sum_{i=1}^{3}k_i\Big)\Big(\TI(k_{1})\overline{\w}(k_{2})\overline{\w}(k_{3})\,-\w(k_{1})\overline{\TI}(k_{2})\overline{\w}(k_{3})+\w(k_{1})\w(k_{2})\TI(k_3)\Big)
\\&=\,i \lambda(2 \pi)^4\delta^4\Big(\sum_{i=1}^{3}k_i\Big)(4 \pi^3)	\prod_{i}\delta(k_i^2+m^2)\theta(k_1^0)\theta(-k_3^0)\Big(\theta(-k_2^0)+\theta(k_2^0)-1\Big)
\\&=\,0.
\end{split}
\ee
Hence, as expected, the 3-point contact interaction does not contribute to the 3-point correlator.
\subsubsection{4-point correlator \label{appsubsecfourpt}}
The four-point on-shell Wightman correlator ${{W}_{\spatinf}(k_1,k_2,k_3,k_4) }$ can be easily computed using its relation to the real part of the S matrix. Here, we show how this computation can be performed directly using Wightman perturbation theory.  To compute the $4-$ point correlator, we need to consider the collapsed contour given in figure \ref{4 point}. 

\paragraph{Contact diagram.}
First consider the correlator at $\Or[g]$ in the presence of an interaction  $\frac{g}{4!}\phi^4 $. At this order, only contact diagrams contribute and we expect them to sum to zero. This can be verified explicitly.

The answer we get from the collapsed contour is as follows,
\be
\begin{split}
	&{W}_{\spatinf\text{contact}}(k_1,k_2,k_3,k_4) 
	\\&
	=\,i g(2 \pi)^4\delta^4\Big(\sum_{i=1}^{4}k_i\Big)\Big({{\ATI}(k_1)}{\wb(k_2)}{\wb(k_3)}{\wb(k_4)}
	-\w(k_1)\TI(k_2){\wb(k_3)}{\wb(k_4)}
	\\&\hspace{5cm}
	+\w(k_1)\w(k_2){\ATI(k_3)}{\wb(k_4)}-\w(k_1)\w(k_2)\w(k_3)\TI(k_4)\Big)
	\\&=(i g)(2 \pi)^4\delta^4\Big(\sum_{i=1}^{4}k_i\Big)	\Big(\prod_{i}(2 \pi)\delta(k_i^2+m^2)\Big)\theta(k_1^0)\theta(-k_4^0)
		\\&\hspace{4.5cm}
	\times\Big({1 \over 2}\theta(-k_3^0)-{1 \over 2}\theta(-k_2^0)\theta(-k_3^0)-{1 \over 2} \theta(k_2^0)+{1 \over 2} \theta(k_2^0)\theta(k_3^0)\Big) \\
&= 0.
\end{split}
\ee
In the last line, we noted that the combination of theta functions answer vanishes for all values of $k_2$ and $k_3$. Hence, as expected, contact diagrams do not contribute to the correlator.
\paragraph{Exchange diagram.}
Let us try to evaluate the $4$-point exchange interaction to $\Or[\lambda^2]$ with the interaction vertices, $\frac{\lambda}{3!}\phi^3$.  The final answer will be a sum of $s$, $t$ and $u$ channel exchange diagrams. 

To evaluate the answer we need to consider 4 different possibilities for $k_2^0$ and $k_3^0$. But, we see that the only nonzero cases correspond to $k_2^0 > 0 , k_3^0<0$ \& $k_2^0 < 0 , k_3^0>0$~. 
As we already have $k_1^0>0$ and $k_4^0<0$, it is impossible to otherwise impose energy-momentum conservation and put all external lines on shell. The final answer is
 \be
 	\label{4exchapp}
 \begin{split}
 &	{W}_{\spatinf\text{exchange}}(k_1,k_2,k_3,k_4)\\ =& \frac{\lambda^{2}}{4}(2 \pi)^4\delta^4\Big(\sum_{i=1}^{4}k_i\Big)\prod_{i}(2 \pi)\delta(k_i^2+m^2)\theta(k_1^0)\theta(-k_4^0) \Big[-\theta(-k_2^0)\theta(k_3^0) \overline{W}(k_1 + k_3) \\
 	&\hspace{2cm}+ \theta(k_2^0)\theta(-k_3^0)(W(k_1 + k_3)- \overline{W}(k_1 + k_3)-W(k_1 + k_2))\Big].
 \end{split}
 \ee
The algebra is straightforward but lengthy; the reader will find details in the associated Mathematica file. As we can see, the final answer is in terms of delta functions of the internal momentum. But the delta functions can not be simultaneously satisfied at the three point vertices.  Hence, the $4$-point exchange diagram vanishes.

On the other hand, it is possible to consider the correlator in the presence of an interaction ${1 \over 2} \lambda \phi^2 \chi$ where $\chi$ is a heavy field with mass $M > 2 m$. The answer has the same form as above with an internal Wightman propagator for $\chi$ instead of $\phi$. In this case, the delta functions can be simultaneously satisfied at each vertex and we could get a non-zero answer. 

\paragraph{$1$-loop diagram.}
We now turn to the $1$- loop diagram to the four-point correlator with the interaction Hamiltonian $ \frac{g}{4!}\phi^4$. It will turn out that only the $s$-channel term contributes to the on-shell correlator. This term is
\be
	\begin{split}
	W_{\spatinf\text{loop}}(k_1,k_2,k_3,k_4)\,=&\,\frac{g}{2}^{2}(2 \pi)^4 \delta^4\Big(\sum_{i=1}^{4}k_i\Big)	\Big(\prod_{i}(2 \pi)\delta(k_i^2+m^2)\Big)\theta(k_1^0)\theta(-k_4^0)\int \frac{d^4k}{(2 \pi)^4}
	\\
	\Big(&{T}(k){T}(k_1+k_2- k )\epsilon_{T }({k_2,k_3})+ \overline{{T}}(k)\overline{{T}}(k_1+k_2- k )\epsilon_{\overline{T}}({k_2,k_3})
	\\
	+&{{W}} (k){{W}}(k_1+k_2-k)\epsilon_{{W}} (k_2,k_3)+\overline{W} (k)\overline{W}(k_1+k_2-k)\epsilon_{\overline{W}}(k_2,k_3)\Big),
\end{split}
\ee 
where, $\epsilon(k_2,k_3)$'s depend on the sign of $k_2^0$ and $k_3^0$ as given in  Table \ref{epsilon_values}. The factor $\frac{1}{2}$ is the symmetry factor for the $1$- loop diagram. 
\begin{table}[H]
	\centering
	\setlength{\extrarowheight}{2pt}
	\begin{tabular}{|c|c|c|}
		\hline
		$\epsilon(k_2,k_3)$&Condition&Value
		\\[0.5ex]
		\hline
		$\epsilon_{T }({k_2,k_3})=\epsilon_{\overline{T}}({k_2,k_3})$ 
		&$k_2^0>0, k^0_3<0$ &  $-\frac{1}{2}$ \\[0.5ex]
		&$k_2^0<0, k^0_3>0$ &  $\,0$ \\[0.5ex]
		\hline
		$\epsilon_{W}({k_2,k_3})$
		&$k_2^0>0, k^0_3<0$ &  $+\frac{1}{4}$ \\[0.5ex]
		&$k_2^0<0, k^0_3>0$ &  $\,0$ \\[0.5ex]
		\hline
		$\epsilon_{\overline{W}}({k_2,k_3})$ 
		&$k_2^0>0, k^0_3<0$ &  $+\frac{1}{2}$ \\[0.5ex]
		&$k_2^0<0, k^0_3>0$ &  $\,0$  \\[0.5ex]
		\hline
	\end{tabular}
		\caption{$\epsilon(k_2,k_3) $ values}
			\label{epsilon_values}
\end{table}
As we discussed, the only possible cases are when $k_2^0 > 0 , k_3^0<0$ \& $k_2^0 < 0 , k_3^0>0$~. Now, for $k_2^0<0, k_3^0>0$, all coefficients are zero. So we only need to look at the non-zero parts coming from $k_2^0>0, k_3^0<0$. 

We can evaluate the integral that involves a product of time-ordered propagators using textbook techniques. By introducing Feynman parameters  and then Wick rotating $k^0= i k_E^0$, we arrive at
\be
\begin{split}
	\Pi_T \equiv &\int \frac{d^4 k}{(2 \pi)^4}  {T}(k){T}(k_1+k_2 -k) 
\\&= -i \int_0 ^1 d x\frac{1}{16 \pi^2} \Big( \frac{2}{ \epsilon} - \gamma+ \log 4 \pi - \log \big( x(1-x)(k_1+k_2)^2+m^2- i \epsilon\big)\Big).
\end{split}
\ee
For the term with anti-time-ordered propagators, one must do the opposite Wick rotation $k^0= - i k_E^0$ to get
\be
\begin{split}
	\Pi_{\overline{T}} &\equiv \int \frac{d^4 k}{(2 \pi)^4}  \overline{{T}}(k)\overline{{T}}(k_1+k_2 - k) \\ &= i \int_0 ^1 dx\frac{1}{16 \pi^2} \Big( \frac{2}{ \epsilon} - \gamma+ \log 4 \pi - \log \big( x(1-x)(k_1+k_2)^2+m^2 + i \epsilon\big)\Big).
\end{split}
\ee
We have regulated both integrals using dimensional regularization although the final answer will be UV-finite.

Note that the imaginary part of these time-ordered and anti-time-ordered terms perfectly cancel each other, while the real parts are same and they add up. The real part can be evaluated by analysing the above expression at the branch cut which starts from $(k_1+k_2)^2<-4 m ^2$. We can check from the following expression that the branch cut contributes as $k_2^0>0$.
\be
	\label{s_chan}
	(k_1+k_2)^2= -2m^2 + 2\Big(-\sqrt{\abs{\vec{k_1}}^2+m^2}\sqrt{\abs{\vec{k_2}}^2+m^2}+\abs{\vec{k_1}}\abs{\vec{k_2}}\cos \theta_{12}\Big)\leq - 4m^2.
\ee
 Integrating the branch-cut contribution we get
\be\label{integral}
	\begin{split}
		\Re[\Pi_{{T}}]=\Re[\Pi_{\overline{T}}] = \frac{1}{16 \pi} \int_{\frac{1}{2}- \frac{1}{2}\sqrt{1+ \frac{4 m^2}{(k_1+k_2)^2}}}^{\frac{1}{2}+ \frac{1}{2}\sqrt{1+ \frac{4 m^2}{(k_1+k_2)^2}}} dx \, = \frac{1}{16 \pi} \sqrt{1+ \frac{4 m^2}{(k_1+k_2)^2}}.
\end{split}
\ee
Now we can compute the integral over $W(k)W(k_1+k_2-k)$
as follows,
\be\label{ww_integral}
	\begin{split}
		&	\int \frac{d^4 k}{(2 \pi)^4} W(k)W(k_1+k_2-k)\\
		= & \int \frac{d^4 k}{(2 \pi)^4} (2 \pi) \theta(k^0)(2 \pi )\theta({p}^0-k^0)\delta(k^2+m^2)\delta(({p}-k)^2+m^2)\qquad \text{here }{p}= k_1+k_2;\\
		= &  \int \frac{d^4 k}{(2 \pi)^4} \times \Big( \int \frac{d^4 k'}{(2 \pi)^4}\, (2 \pi)^4 \delta^4 ({p}- k - k' ) \Big)\times  (2 \pi) \theta(k^0) (2 \pi) \theta(k'^0)\delta(k^2+m^2)\delta(k'^2+m^2)\\
		= & \int \frac{d^3 \mathbf{k}}{(2 \pi)^3} \frac{1}{2 \sqrt{\mathbf{k}^2+m^2}}\frac{d^3 \mathbf{k'}}{(2 \pi)^3} \frac{1}{2\sqrt{\mathbf{k'}^2+m^2}} (2 \pi)^4\delta(p^0 - \sqrt{\mathbf{k}^2+m^2}- \sqrt{\mathbf{k'}^2+m^2})\delta^3(\mathbf{p}- \mathbf{k}- \mathbf{k'}) \\
		= & \frac{1}{8 \pi} \sqrt{1+ 4\frac{m^2}{(k_1+k_2)^2}}.
	\end{split}
\ee
Let us now consider the integral over $\overline{W}(k)\overline{W}(k_1+k_2-k)$. Let us choose a frame, $	k_1+k_2= \{ q,0,0,0\}$ where $\vec{k}_1+\vec{k}_2=0$. As $k_1^0>0, k_2^0>0$ we get $q>0$. Then the delta functions, $\delta(k^2+m^2)$ and $\delta((k_1+k_2-k)^2+m^2)$  would imply that $k^{0}$ must be greater than zero. Thus the integral $	\int \frac{d^4 k}{(2 \pi)^4}\overline{W} (k)\overline{W}(k_1+k_2-k)$ does not contribute.
 
The final non-zero result for $k_2^0>0 ,k_3^0<0$ (contributions from $\Re[TT]+ \Re[\overline{T}\overline{T}]+ W W$) is
\be
	\label{loop_final}
	\begin{split}
		&W_{\spatinf\text{loop}}(k_1,k_2,k_3,k_4)\\
		\,= &- \frac{g^2}{64 \pi} (2 \pi)^8 \delta^4\Big(\sum_{i=1}^{4}k_i\Big)	\prod_{i}\delta(k_i^2+m^2)\theta(k_1^0)\theta(k_2^0)\theta(-k_3^0)\theta(-k_4^0) \sqrt{1+\frac{4 m^2}{(k_1+k_2)^2}}.
	\end{split}
\ee

In the same way, we can compute the $t$ and $u$ channel diagrams. These diagrams do not contribute to the answer as we now explain.

To start with, the answer obtained by evaluating the diagrams is similar to the s-channel answer except that the coefficients of the different terms change according to the tables \ref{epsilon_values2} and \ref{epsilon_values3}.
\begin{table}[H]
	\centering
	\setlength{\extrarowheight}{2pt}
	\begin{tabular}{|c|c|c|}
		\hline
		$\epsilon(k_2,k_3)$ for $t$ channel&Condition&Value
		\\[0.5ex]
		\hline
		$\epsilon_{T }({k_2,k_3})=\epsilon_{\overline{T}}({k_2,k_3})$ 
		&$k_2^0>0, k^0_3<0$ &  $-\frac{1}{2}$ \\[0.5ex]
		&$k_2^0<0, k^0_3>0$ &  $\,0$ \\[0.5ex]
		\hline
		$\epsilon_{W }({k_2,k_3})$
		&$k_2^0>0, k^0_3<0$ &  $+\frac{3}{4}$ \\[0.5ex]
		&$k_2^0<0, k^0_3>0$ &  $\,0$ \\[0.5ex]
		\hline
		$\epsilon_{\overline{W} }({k_2,k_3})$ 
		&$k_2^0>0, k^0_3<0$ &  $+\frac{1}{4}$ \\[0.5ex]
		&$k_2^0<0, k^0_3>0$ &  $\,- \frac{1}{4}$  \\[0.5ex]
		\hline
	\end{tabular}
	\caption{$\epsilon(k_2,k_3) $ values for $t$ channel}
	\label{epsilon_values2}
\end{table}

\begin{table}[H]
	\centering
	\setlength{\extrarowheight}{2pt}
	\begin{tabular}{|c|c|c|}
		\hline
		$\epsilon(k_2,k_3)$ for $u$ channel&Condition&Value
		\\[0.5ex]
		\hline
		$\epsilon_{T }({k_2,k_3})=\epsilon_{\overline{T}}({k_2,k_3})$ 
		&$k_2^0>0, k^0_3<0$ &  $-\frac{1}{2}$ \\[0.5ex]
		&$k_2^0<0, k^0_3>0$ &  $\,0$ \\[0.5ex]
		\hline
		$\epsilon_{W}({k_2,k_3})$
		&$k_2^0>0, k^0_3<0$ &  $+\frac{1}{2}$ \\[0.5ex]
		&$k_2^0<0, k^0_3>0$ &  $\,0$ \\[0.5ex]
		\hline
		$\epsilon_{\overline{W}}({k_2,k_3})$ 
		&$k_2^0>0, k^0_3<0$ &  $+\frac{1}{2}$ \\[0.5ex]
		&$k_2^0<0, k^0_3>0$ &  $\,0$  \\[0.5ex]
		\hline
	\end{tabular}
	\caption{$\epsilon(k_2,k_3) $ values for $u$ channel}
	\label{epsilon_values3}
\end{table}
\textbf{t- channel:}
In this case $T(k)T(k_1+k_3-k)$ and $\overline{T}(k)\overline{T}(k_1+k_3-k)$ terms contribute only when $k_2^0>0, k_3^0<0$. The imaginary parts cancel with each other as before. The real parts could contribute.  However, we notice that to get a nonzero real part, we should have $(k_1+k_3)^2< -4m^2$. But
\be
	\label{t_chan}
	(k_1+k_3)^2= -2m^2 + 2\Big(\sqrt{\abs{\vec{k_1}}^2+m^2}\sqrt{\abs{\vec{k_3}}^2+m^2}+\abs{\vec{k_1}}\abs{\vec{k_3}}\cos \theta_{13}\Big)\geq 0.
\ee
Hence the real parts also do not contribute.

Consider the case when, $k_{2}^0<0, k_3^0>0$. We can now choose a frame $k_1+k_3= \{ q,0,0,0\}$, where $q>0$. (All the expressions are Lorentz-invariant and so if they vanish in one frame they vanish in all frames.)  We see that the only nonzero contribution comes from the integral over $\overline{W} (k)\overline{W}(k_1+k_3-k)$. But this requires $k^0 < 0$ and also that $k_1 + k_3 - k$ is on-shell. This is impossible and hence this term does not contribute. For the case, $k_{2}^0>0, k_3^0<0$, we can go to a frame where $k_1^0+k_3^0=0$. In this frame, it is clear that the integral over both $W(k)W(k_1+k_3-k)$ and $\overline{W} (k)\overline{W}(k_1+k_3-k)$ vanishes.

\textbf{u- channel:} Following the same analysis as the $t$ channel we find that $T(k)T(k_1+k_4-k)$ and $\overline{T}(k)\overline{T}(k_1+k_4-k)$ do not contribute. In this case, we are given that $k_1^0 > 0$ and $k_4^0 < 0$ and since both momenta are on shell we can choose a frame where $k_1^0+k_4^0=0$. It is then clear that the integral over both ${W} (k){W}(k_1+k_4-k)$ and $\overline{W} (k)\overline{W}(k_1+k_4-k)$ vanishes.

\subsubsection{$5$-point correlator}
We want to compute the correlator  ${{W}_{\spatinf}(k_1,k_2,k_3,k_4,k_5) }$, with, $k_1^0>0, k_5^0<0$. To compute this correlator, we need to consider the collapsed contour shown in the following figure.
\begin{figure}[H]
	\centering
	\hspace{1.7cm}
	\includegraphics[scale=0.3]{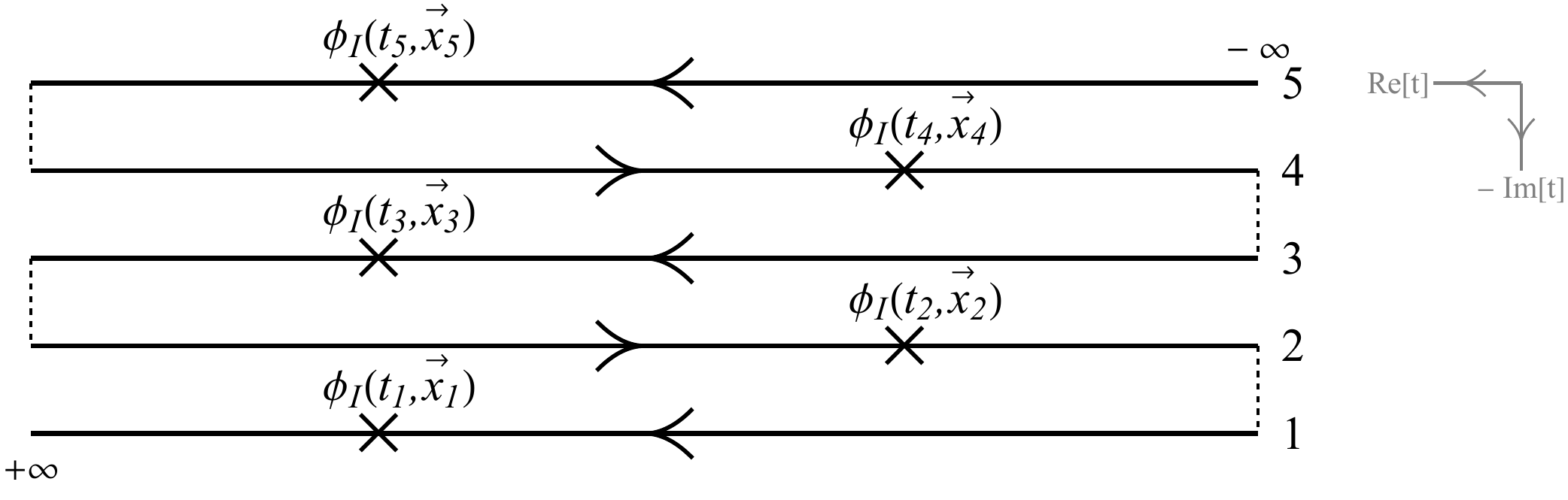}
	\caption{Collapsed contour lines for 5 point function}
	\label{5point contour}
\end{figure}

We start by studying the interaction Hamiltonian $\frac{g}{4!}\phi^{4} + \frac{\lambda}{3!}\phi^{3}$. The manipulations are tedious and are performed in the associated Mathematica file. We get the answer in terms of the delta functions of the exchange momentum as displayed in section \ref{fivepointvac}. But these delta functions can never be satisfied consistent with energy-momentum conservation when the external legs are on shell. Therefore the $5$-point correlator vanishes at $\Or[\lambda g]$. 

However, it is possible to consider the another theory with the  interaction Hamiltonian as $\frac{\tilde{\lambda}}{2!}\phi^2 \chi+\frac{\tilde{g}}{3!}\phi^3\chi$, where, $\chi$ is a scalar field with heavy mass  $M > 2 m$. The intermediate Wightman and anti-Wightman propagators for $\phi$ are replaced with propagators for $\chi$ and these delta functions can yield a nonzero answer. The final answer is as given in equation \eqref{5 pnt vacuum}.
\subsubsection{$6-$point correlator}
\begin{figure}[H]
	\centering
	\hspace{1.7cm}
	\includegraphics[scale=0.3]{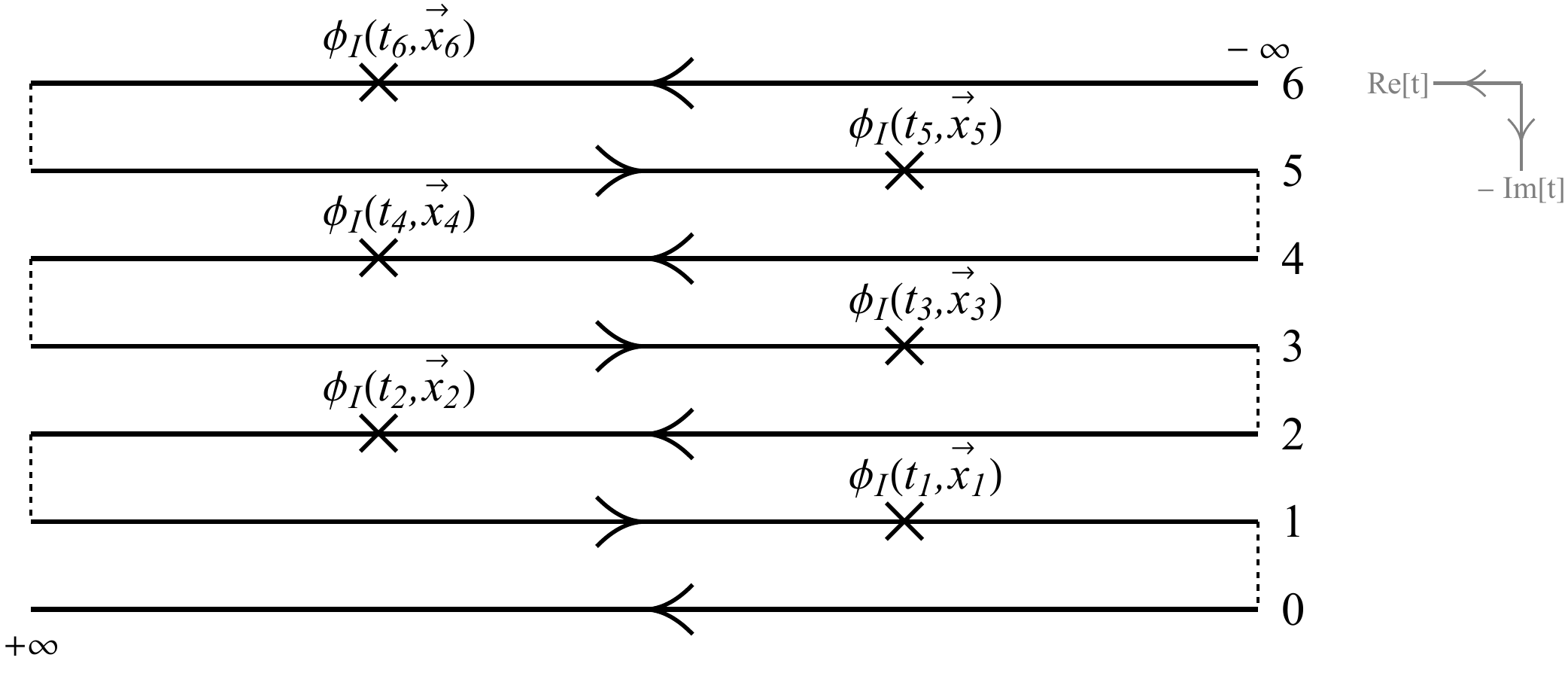}
	\caption{Collapsed contour lines for 6 point function}
	\label{6point contour}
\end{figure}

To compute the  $6-$point correlator  ${{W}_{\spatinf}(k_1,k_2,k_3,k_4,k_5,k_6) }$  we need to consider the contour given in figure \ref{6point contour}.
For a $6-$ point exchange diagram with the interaction Hamiltonian $\frac{g^4}{4!}\phi^4$, we get the answer given in equation \eqref{6 pnt vacuum}. Here too,  the final answer is in terms of the delta functions of the exchange momentum. However, this combination of delta functions can yield a nonzero answer.
\subsection{Wightman functions of points at $\hat{i}^0$ in non-vacuum state}
In this section, we will evaluate correlators $W^{\Psi_1, \Psi_2}_{\spatinf}(k_1,\ldots,k_n)$ with arbitrary external states $\Psi_1$ and $\Psi_2$. When the external points are at $i^{+}/i^{-}$, we use the rules given in \eqref{rules_non-vacuum} and when they are at ${\spatinf}$, we use the rules \eqref{modifiedrules}.
\subsubsection{$4-$point exchange diagram}
\begin{figure}[H]
	\centering
	\hspace{1.7cm}
	\includegraphics[scale=0.3]{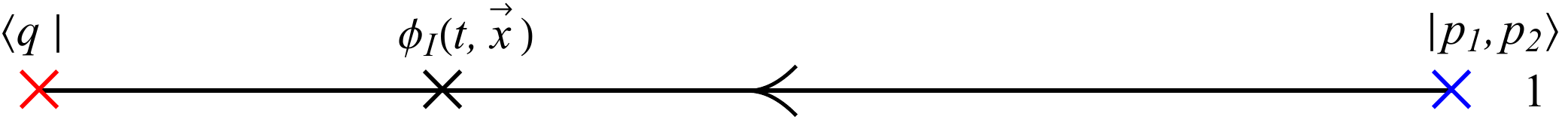}
	\caption{Collapsed contour lines for 4 point function in non-vacuum state}
	\label{4point_122}
\end{figure}

Let us  evaluate the $4-$ point exchange diagram $W^{\vec{p}_1\vec{p}_2, \vec{q}}_{\spatinf}(k)$ in $\lambda\phi^3$ theory, where $\vec{p}_1,\vec{p}_2$  are at ${i^-}$, $\vec{q}$ is at $i^{+}$ and $k$ is at $\spatinf$.
We need to consider the contour given in figure 
\ref{4point_122}. 

  We get the following answer from the contour after applying the rules \eqref{rules_non-vacuum} and \eqref{modifiedrules}.
\be
\begin{split}
&W_{\spatinf}^{\vec{q},\vec{p}_1\vec{p}_2}(k)\\
=&\,-\frac{\lambda^{2}}{2}(2 \pi)^4\delta^4\Big(q+k+\sum_{i=1}^{2}p_i\Big)(2 \pi)	\delta(k^2+m^2)\theta(k^0)
\Big( T(q + k)+ T(q+p_1)+T(q+p_2)\Big).
\end{split}
\ee
We can also compute $W_{\spatinf}^{\Omega,\vec{p}_1 \vec{p}_2}(k_1,k_2)$ where the \outip state is the vacuum. However, the action of the field on the left $\phi_{\spatinf}(k_1)$ creates a single-particle state 
\be
\langle \Omega | \phi_{\spatinf}(k_1) = (2 \pi) \delta(k_1^2 + m^2) \theta(k_1^0) \langle \vec{k}_1|.
\ee
Therefore 
\be
W_{\spatinf}^{\Omega,\vec{p}_1 \vec{p}_2}(k_1,k_2) = (2 \pi) \delta(k_1^2 + m^2) \theta(k_1^0) W_{\spatinf}^{\vec{k}_1, \vec{p}_1 \vec{p}_2}(k_2).
\ee
\subsubsection{$6-$ point exchange diagram}
\begin{figure}[H]
	\centering
	\hspace{1.7cm}
	\includegraphics[scale=0.3]{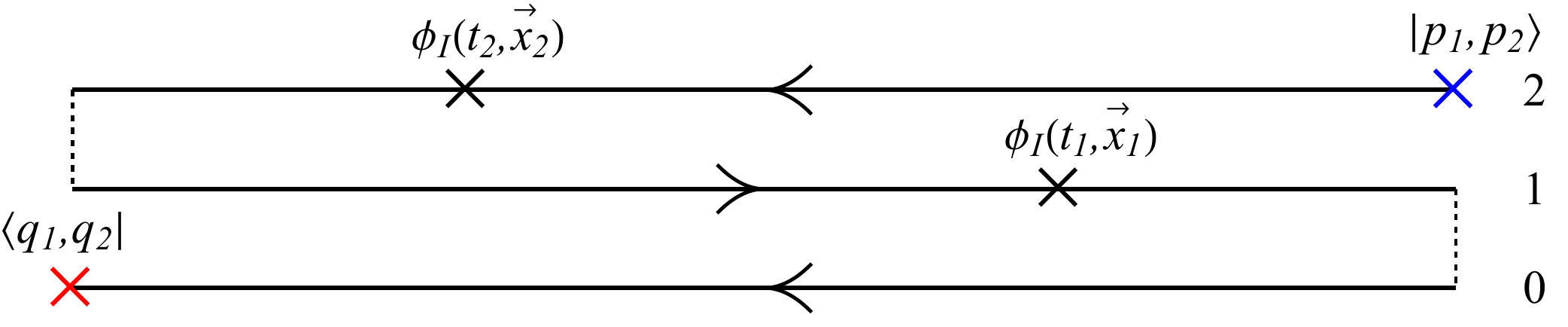}
	\caption{Collapsed contour lines for 6 point function in non-vacuum state}
	\label{6point_222}
\end{figure}

Finally, we consider the correlator with two insertions each at $\spatinf$, $i^+$, $i^-$: $W^{\vec{q}_1 \vec{q}_2, \vec{p}_1 \vec{p}_2}_{\spatinf}(k_1,k_2)$. Here $\vec{q}_1,\vec{q}_2$ are at $i^+$, $\vec{p}_1,\vec{p}_2$ are at $i^-$ and $k_1,k_2$ are at $\spatinf$. In a theory with a ${g \phi^4 \over 4!}$ interaction, this receives contributions from exchange diagrams.   We need to consider the contour given in Figure \ref{6point_222}. 

After some algebra, we get the answer given in \eqref{6 pnt nonvacuum}.

\bibliographystyle{utphys}
\bibliography{references}

\end{document}